\documentclass[a4paper]{article}

\pdfoutput=1

\usepackage{amssymb,amsmath,latexsym,amsthm,amsfonts}
\usepackage{graphicx}%,showlabels}
\usepackage{fancyhdr,enumerate,scrtime}
\usepackage{fullpage}

\usepackage{color}
\usepackage{bbm}
\usepackage{mathrsfs}
\usepackage{ifthen}
\usepackage{natbib}

\bibpunct{(}{)}{,}{a}{}{,}

\newtheorem{example}{Example}
\newtheorem{example*}{Example}

\newcommand{\bx}{\mathbf{x}}
\newcommand{\bz}{\mathbf{z}}
\newcommand{\dd}{\text{d}}
\newcommand{\ds}{\displaystyle}
\newcommand{\bn}{\mathbf{n}}

\definecolor{LightGrey}{rgb}{0.86,0.75,0.43}

\title{Bayesian Inference on Mixtures of Distributions
\footnote{Kate Lee is a PhD candidate at the Queensland University
of Technology, Jean-Michel Marin is a researcher at INRIA,
Universit\'e Paris Sud, and adjunct professor at
\'Ecole Polytechnique, Kerrie Mengersen is professor at the
Queensland University of Technology, and Christian P. Robert is professor
in Universit\'e Paris Dauphine and head of the Statistics
Laboratory of CREST.}}

\author{Kate Lee \\ Queensland University of Technology \and
Jean-Michel Marin \\ INRIA Saclay, Projet \textsc{select},
Universit\'e Paris-Sud and CREST, INSEE \and
Kerrie Mengersen \\ Queensland University of Technology \and
Christian Robert \\ Universit\'e Paris Dauphine and CREST, INSEE
}

\begin{document}

\maketitle

\begin{abstract}
This survey covers state-of-the-art Bayesian techniques for the estimation
of mixtures. It complements the earlier \cite{marin:mengersen:robert:2005}
by studying new types of distributions, the multinomial, latent class
and $t$ distributions. It also exhibits closed form solutions for Bayesian
inference in some discrete setups. Lastly, it sheds a new light on the
computation of Bayes factors via the approximation of \cite{chib:1995}.
\end{abstract}

% \pagenumbering{arabic}
% \chapter{Bayesian Inference on Mixtures of Distributions}
% \chapterauthors{Kate Lee, Jean-Michel Marin, Kerrie Mengersen and Christian P. Robert}

%%%%%%%%%%%%%%%%%%%%%%%%%%%%%%%%%%%%%%%
\section{Introduction}\label{sec:intro}
%%%%%%%%%%%%%%%%%%%%%%%%%%%%%%%%%%%%%%%

Mixture models are fascinating objects in that, while based on elementary
distributions, they offer a much wider range of modeling possibilities than their components.
They also face both highly complex computational challenges and delicate 
inferential derivations.  Many statistical advances have stemmed from their study, 
the most spectacular example being the EM algorithm. In this short review,
we choose  to focus solely on the Bayesian approach to those models \citep{robert:casella:2004}.
\cite{fruhwirth:2006} provides a book-long and in-depth coverage of the Bayesian processing of mixtures,
to which we refer the reader whose interest is woken by this short review, while \cite{maclachlan:peel:2000}
give a broader  perspective.

Without opening a new debate about the relevance of the Bayesian approach in general, we note that
the Bayesian paradigm \citep[see, e.g.,][]{robert:2001} allows for probability statements to be made directly 
about the unknown parameters of a mixture model, and for prior or expert opinion to be included in the analysis. In 
addition, the latent structure that facilitates the description of a mixture model can be naturally 
aggregated with the unknown parameters (even though latent variables are {\em not} parameters) 
and a global posterior distribution can be used to draw inference
about both aspects at once. 

This survey thus aims to introduce the reader to the construction, prior modelling, estimation 
and evaluation of mixture distributions within a Bayesian paradigm. Focus is on both Bayesian inference
and computational techniques, with light shed on the implementation of the most common samplers. We also
show that exact inference (with no Monte Carlo approximation) is achievable in some particular settings
and this leads to an interesting benchmark for testing computational methods.

\newpage

In Section \ref{sec:finite}, we introduce mixture models, including the missing data structure
that originally appeared as an essential component of a Bayesian analysis,
along with the precise derivation of the exact posterior
distribution in the case of a mixture of Multinomial distributions. Section \ref{sec:conundrum}
points out the fundamental difficulty in conducting Bayesian inference with such objects, along with a discussion
about prior modelling. Section \ref{sec:MCMC} describes the appropriate
MCMC algorithms that can be used for the approximation to the posterior
distribution on mixture parameters, followed by an extension of this analysis in Section \ref{sec:mocho} 
to the case in which the number of components is unknown and may be derived from approximations to
Bayes factors, including the technique of \cite{chib:1995} and the robustification of
\cite{berkhof:vanmechelen:gelman:2003}.

%%%%%%%%%%%%%%%%%%%%%%%%%%%%%%%%%%%%%%%%%%%%%%%%%%%%%%%%
\section{Finite mixtures}\label{sec:finite}
%%%%%%%%%%%%%%%%%%%%%%%%%%%%%%%%%%%%%%%%%%%%%%%%%%%%%%%%

\subsection{Definition}\label{sub:mix.def}
A mixture of distributions is defined as a convex combination
$$
\sum_{j=1}^J p_j f_j(x)\,,\quad \sum_{j=1}^J p_j =1\,,\quad p_j>0\,,\quad J>1\,,
$$
of standard distributions $f_j$. The $p_j$'s are called {\em weights} and
are most often unknown. In most cases, the interest is in having the $f_j$'s
parameterised, each with an unknown parameter $\theta_j$, leading to the 
generic parametric mixture model
\begin{equation}\label{eq:mix}
\sum_{j=1}^J p_j f(x|\theta_j)\,.
\end{equation}

The dominating measure for (\ref{eq:mix}) is arbitrary and therefore the nature of the mixture observations
widely varies. For instance, if the dominating measure is the counting measure on the simplex of $\mathbb{R}^m$
$$
\mathcal{S}_{m,\ell}=\left\{ (x_1,\ldots,x_m); \sum_{i=1}^m x_i = \ell \right\}\,,
$$
the $f_j$'s may be the product of $\ell$ independent Multinomial distributions, denoted
``$\mathcal{M}_m(\ell; q_{j1},...,q_{jm})=\otimes_{i=1}^\ell \mathcal{M}_m(1; q_{j1},...,q_{jm})$'',
with $m$ modalities, and the resulting mixture
\begin{equation}
\label{eq:multino}
\sum_{j=1}^J p_j \mathcal{M}_m(\ell;q_{j1},\ldots,q_{jm})
\end{equation}
is then a possible model for repeated observations taking place in $\mathcal{S}_{m,\ell}$. Practical 
occurrences of such models are repeated observations of {\em contingency tables}.
% with extra-binomial variation: if we observe 
% $$
% x_{ij}, \quad i=1,\ldots,m,\,j=1,\ldots,n\,,
% $$
% the variation observed on one component, $x_{1j}$ say, may well exceed the variation 
% expected from a binomial distribution. 
In situations when contingency tables tend to vary more than expected, a mixture of Multinomial 
distributions should be more appropriate than a single Multinomial distribution and
it may also contribute to separation of the observed tables in homogeneous classes 
% (i.e., in groups where the $q_{ji}$'s are close enough to be considered as identical in $i$).
In the following, we note $q_{j\cdot}=(q_{j1},\ldots,q_{jm})$.

\begin{figure}
\centerline{\includegraphics[height=7truecm]{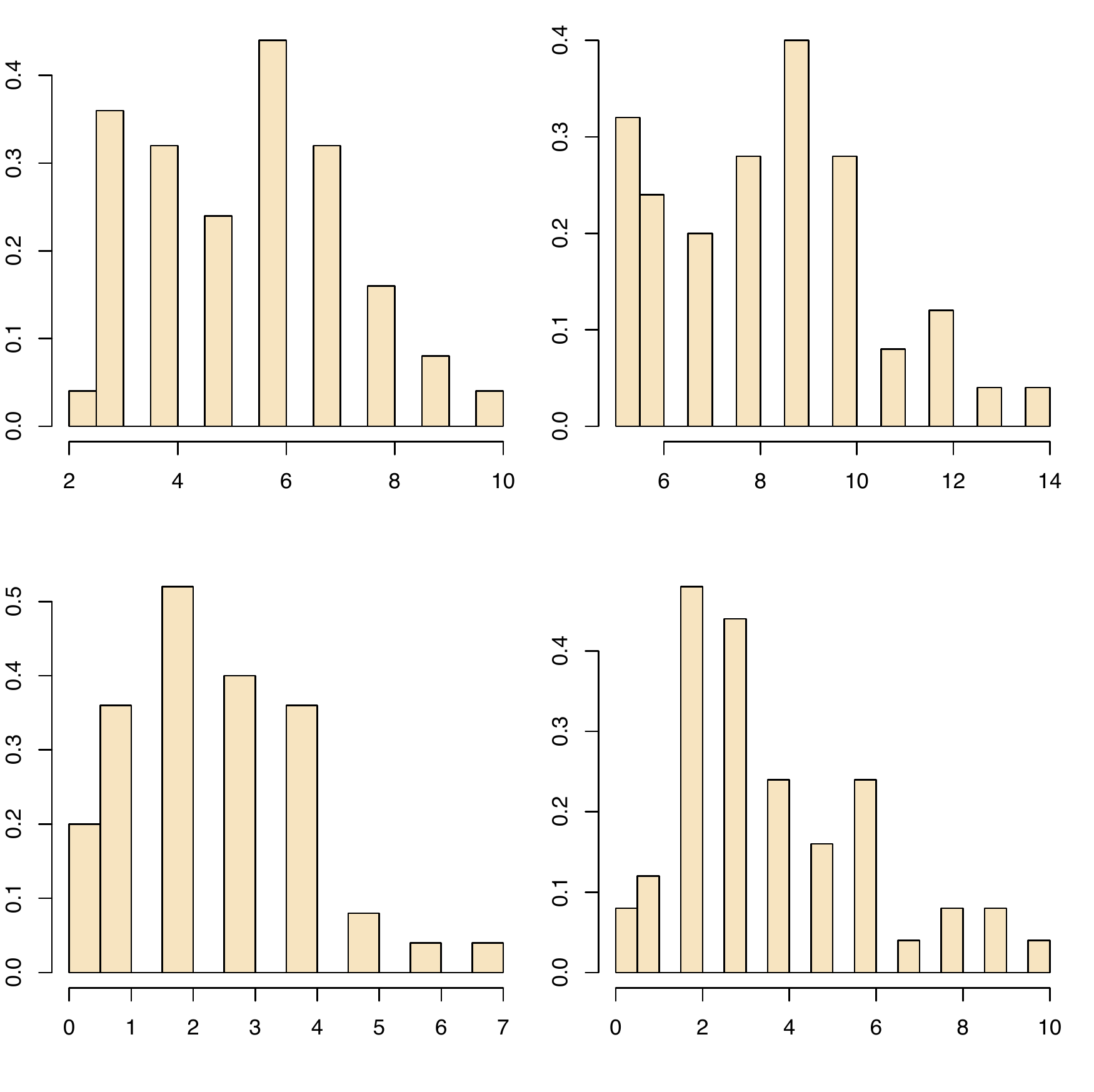}}
\caption{\label{fig:multist}
For $J=2$, $p_1=p_2=.5$, $q_{1\cdot}=(.2,.5,.2,.1)$, $q_{2\cdot}=(.3,.3,.1,.3)$,
$\ell=20$ and $n=50$ independent simulations: histograms of the $m=4$ entries.
}
\end{figure}

\begin{example}
\begin{em}
For $J=2$, $m=4$, $p_1=p_2=.5$, $q_{1\cdot}=(.2,.5,.2,.1)$, $q_{2\cdot}=(.3,.3,.1,.3)$ and 
$\ell=20$, we simulate $n=50$ independent realisations from model (\ref{eq:multino}). That corresponds to simiulate some $2\times 2$ contingency tables whose total 
sum is equal to $20$. Figure \ref{fig:multist} gives the histograms for the four entries of the contingency tables. \hfill$\blacktriangleleft$
\end{em}
\end{example}

Another case where mixtures of Multinomial distributions occur is the {\em latent class model}
where $d$ discrete variables are observed on each of $n$ individuals \citep{magidson:vermunt:2000}. The observations
	$(1\le i\le n)$ are $\bx_i = (x_{i1},\ldots,x_{id})$, 
with $x_{iv}$ taking values within the $m_v$ modalities of the $v$-th variable. The distribution of $\bx_i$ 
is then 
$$
\sum_{j=1}^J p_j \prod_{i=1}^d \mathcal{M}_{m_i}\left(1;q_1^{ij},\ldots,q_{m_i}^{ij}\right)\,,
$$
so, strictly speaking, this is a mixture of {\em products} of Multinomials. The applications of this 
peculiar modelling are numerous: in medical studies, it can be used to associate several symptoms or 
pathologies; in genetics, it may indicate that the genes corresponding to the variables are not 
sufficient to explain the outcome under study and that an additional (unobserved) gene may be influential.
Lastly, in marketing, variables may correspond to categories of products, modalities to brands, and components
of the mixture to different consumer behaviours: identifying to which group a customer belongs may help in
suggesting sales, as on Web-sale sites.

Similarly, if the dominating measure is the counting measure on the set of the integers $\mathbb{N}$,
the $f_j$'s may be Poisson distributions $\mathcal{P}(\lambda_j)$ $(\lambda_j>0)$. We aim then to make inference
about the parameters $(p_j,\lambda_j)$ from a sequence $(x_i)_{i=1,\ldots,n}$ of integers.

The dominating measure may as well be the Lebesgue measure on $\mathbb{R}$, in which case the
$f(x|\theta)$'s may all be normal distributions or Student's $t$ distributions (or even a mix
of both), with $\theta$ representing the unknown mean and variance, or the unknown mean and variance
and degrees of freedom, respectively. Such a model is appropriate for datasets presenting
multimodal or asymmetric features, like the aerosol dataset from \cite{nilsson:kulmala:2006} 
presented below.

\begin{figure}
\centerline{\includegraphics[height=7truecm]{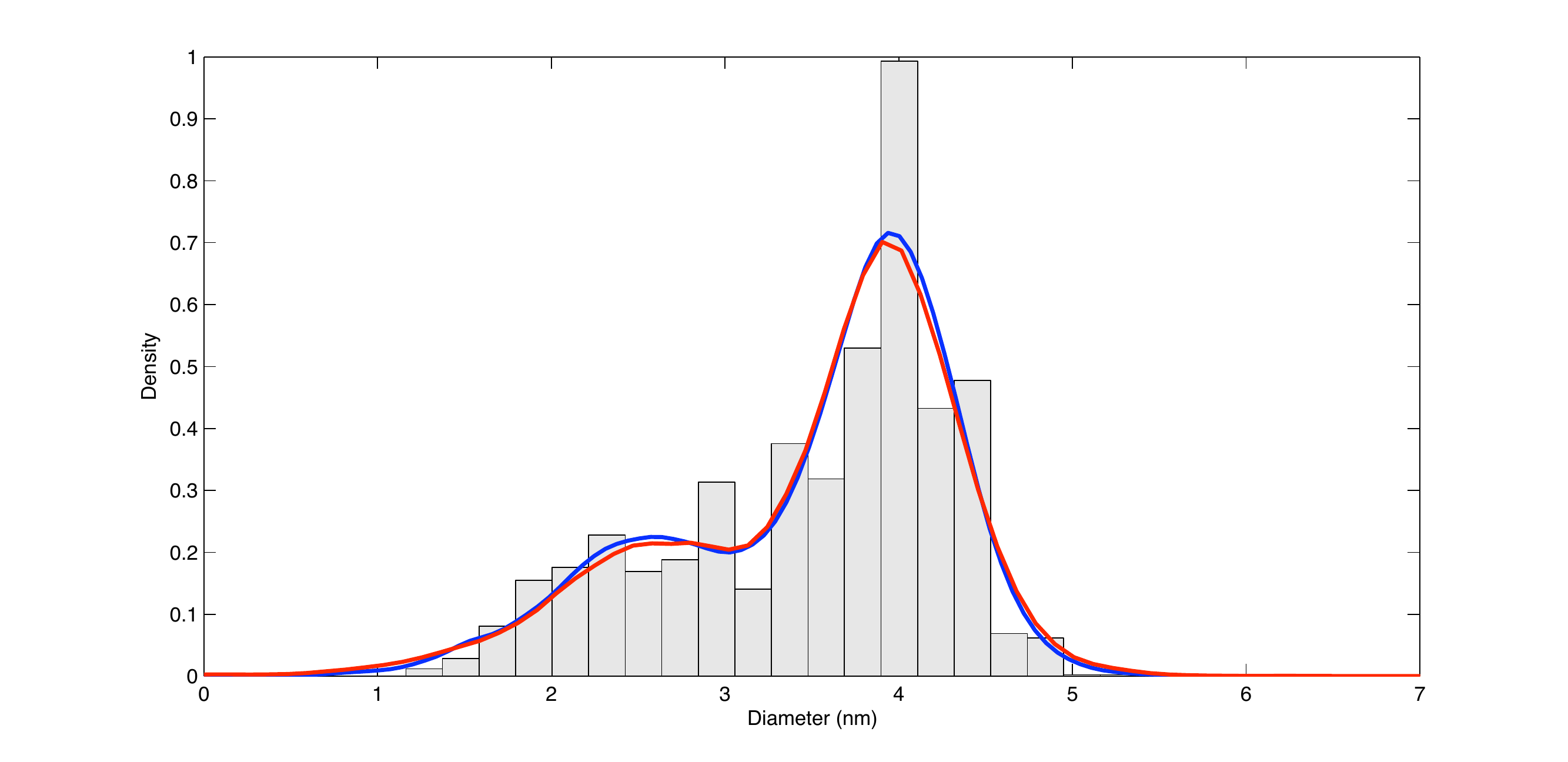}}
\caption{\label{fig:aero1}
Histogram of the aerosol diameter dataset, along with a normal {\em (red)} and a $t$ 
{\em (blue)} modelling.
}
\end{figure}

\begin{example}\label{ex:aero}
\begin{em}
The estimation of particle size distribution is important in understanding the 
aerosol dynamics that govern aerosol formation, which is of interest in environmental and health modelling.
One of the most important physical properties of aerosol particles is their size; the concentration of aerosol particles
in terms of their size is referred to as the {\em particle size distribution.} 

The data studied by \cite{nilsson:kulmala:2006} and represented in 
Figure \ref{fig:aero1} is from Hyyti\"al\"a, a measurement station in 
Southern Finland. It corresponds to a full day of measurement, taken at 
ten minute intervals. \hfill$\blacktriangleleft$
\end{em}
\end{example}

\medskip
While the definition (\ref{eq:mix}) of a mixture model is elementary, its simplicity does not
extend to the derivation of either the maximum likelihood estimator (when it exists) or of Bayes estimators. 
In fact, if we take $n$ iid observations $\mathbf{x}=(x_1,\ldots,x_n)$ from (\ref{eq:mix}), with
parameters 
$$
\mathbf{p}=(p_1\ldots,p_J)\quad\text{and}\quad\boldsymbol{\theta}=(\theta_1,\ldots,\theta_J)\,,
$$ 
the full computation of the posterior distribution and in particular  
the explicit representation of the corresponding posterior expectation involves the expansion of 
the likelihood
\begin{equation}\label{eq:likey}
\text{L}(\boldsymbol{\theta},\mathbf{p}|\mathbf{x}) 
= \prod_{i=1}^n \sum_{j=1}^J p_j f\left(x_i|\theta_j\right) 
\end{equation}
into a sum of $J^n$ terms. with some exceptions (see, for example Section \ref{sec:conundrum}). This is thus
computationally too expensive to be used for more than a few observations. 
This fundamental computational difficulty in dealing with the models (\ref{eq:mix}) explains why those models
have often been at the forefront for applying new technologies (such as MCMC algorithms, see Section \ref{sec:MCMC}).

\subsection{Missing data}\label{sub:mix.miss}
Mixtures of distributions are typical examples of {\em latent variable} (or
{\em missing data}) models in that a sample
$x_1,\ldots,x_n$ from (\ref{eq:mix}) can be seen as a collection of subsamples
originating from each of the $f\left(x_i|\theta_j\right)$'s, when both the size and
the origin of each subsample may be unknown. Thus, each of the $x_i$'s in the sample
is {\em a priori} distributed from any of the $f_j$'s with probabilities $p_j$. Depending on the setting, the 
inferential goal behind this modeling may be to reconstitute the original homogeneous subsamples, sometimes called {\em clusters}, or to
provide estimates of the parameters of the different components, or even to
estimate the number of components. 

The missing data representation of a mixture distribution can be exploited as a technical
device to facilitate (numerical) estimation. By a 
demarginalisation argument, it is always possible to associate to a random variable 
$x_i$ from a mixture \eqref{eq:mix} a second (finite) random variable $z_i$ such that
\begin{equation}\label{eq:latent} 
x_i|z_i=z \sim f(x|\theta_z)\,,\quad \mathbb{P}\left(z_i=j\right)=p_j\,.
\end{equation}
This auxiliary variable $z_i$ identifies to which component the observation $x_i$ belongs.
Depending
on the focus of inference, the $z_i$'s may [or may not] be part of the
quantities to be estimated.
In any case, keeping in mind the availability of such variables helps into drawing
inference about the ``true" parameters. This is the technique behind the EM algorithm of
\cite{dempster:laird:rubin:1977} as well as the ``data augmentation" algorithm of
\cite{tanner:wong:1987} that started MCMC algorithms.

\subsection{The necessary but costly expansion of the likelihood}\label{sub:comics}
As noted above, the likelihood function (\ref{eq:likey}) involves $J^n$ terms when the 
$n$ inner sums are expanded, that is, when all the possible values of the missing variables $z_i$ are 
taken into account. While the likelihood at a given value $\left(\boldsymbol{\theta},\mathbf{p}\right)$
can be computed in $\mathrm{O}(nJ)$ operations,
the computational difficulty in using the expanded version of (\ref{eq:likey}) 
precludes analytic solutions via maximum likelihood or Bayesian inference.
Considering $n$ iid observations from
model \eqref{eq:mix}, if $\pi\left(\boldsymbol{\theta},\mathbf{p}\right)$ denotes the
prior distribution on $\left(\boldsymbol{\theta},\mathbf{p}\right)$, the posterior distribution is 
naturally given by
$$%\begin{equation}\label{eq:post}
\pi\left(\boldsymbol{\theta},\mathbf{p}|\mathbf{x}\right)\propto
\left(\prod_{i=1}^n\sum_{j=1}^J p_jf\left(x_i|\theta_j\right)\right)\pi\left(\boldsymbol{\theta},\mathbf{p}\right).
$$%\end{equation}
It can therefore be computed in $\mathrm{O}(nJ)$ operations
up to the normalising [marginal] constant, but, similar to
the likelihood, it does not provide an intuitive distribution unless expanded.

Relying on the auxiliary variables $\mathbf{z}=\left(z_1,\ldots,z_n\right)$ defined in 
\eqref{eq:latent}, we take $\mathcal{Z}$ to be the set of all $J^n$ allocation vectors $\mathbf{z}$. 
For a given vector $\left(n_1,\ldots,n_J\right)$ of the simplex $\left\{n_1+\ldots+n_J=n\right\}$, we define a subset 
of $\mathcal{Z}$,
$$
\mathcal{Z}_j = \left\{\mathbf{z}:\sum_{i=1}^n\mathbb{I}_{z_i=1}=n_1,
\ldots,\sum_{i=1}^n\mathbb{I}_{z_i=J}=n_J\right\}\,,
$$
that consists of all allocations $\mathbf{z}$ with the given allocation 
sizes $\left(n_1,\ldots,n_J\right)$,
relabelled by $j\in\mathbb{N}$ when using for instance the lexicographical ordering on $\left(n_1,\ldots,n_J\right)$.
The number of nonnegative integer solutions to the decomposition of $n$ into $k$ parts
such that $n_1+\ldots+n_J=n$ is equal to \citep{feller:1970}
$$
r={n+J-1 \choose n} \,.
$$
Thus, we have the partition $\displaystyle \mathcal{Z}=\cup_{j=1}^r \mathcal{Z}_j$. Although the total number
of elements of $\mathcal{Z}$ is the typically unmanageable $J^n$, the number of partition sets is much 
more manageable since it is of order $n^{J-1}/(J-1)!$. It is thus possible to envisage an exhaustive 
exploration of the $\mathcal{Z}_j$'s. (\citealp{Casella:Robert:Wells:1999} did take
advantage of this decomposition to propose a more efficient important sampling approximation to the
posterior distribution.)

The posterior distribution can then be written as
\begin{equation}
\pi\left(\boldsymbol{\theta},\mathbf{p}|\mathbf{x}\right)=\sum_{i=1}^r\sum_{\mathbf{z}
\in\mathcal{Z}_i}\omega\left(\mathbf{z}\right)
\pi\left(\boldsymbol{\theta},\mathbf{p}|\mathbf{x},\mathbf{z}\right) \,,
\end{equation}
where $\omega\left(\mathbf{z}\right)$ represents the posterior probability of the given allocation 
$\mathbf{z}$. (See Section \ref{sec:xact} for a derivation of $\omega\left(\mathbf{z}\right)$.)  
Note that with this representation, a Bayes estimator of $\left(\boldsymbol{\theta},\mathbf{p}\right)$
can be written as
\begin{equation}\label{eq:bigsum}
\sum_{i=1}^r\sum_{\mathbf{z}\in\mathcal{Z}_i}\omega\left(\mathbf{z}\right)
\mathbb{E}^\pi\left[\boldsymbol{\theta},\mathbf{p}|\mathbf{x},\mathbf{z}\right] \,.
\end{equation}
This decomposition makes a lot of sense from an inferential point 
of view: the Bayes posterior distribution simply considers each possible allocation
$\mathbf{z}$ of the dataset, allocates a posterior probability 
$\omega\left(\mathbf{z}\right)$ to this allocation, and then constructs a posterior
distribution $\pi\left(\boldsymbol{\theta},\mathbf{p}|\mathbf{x},\mathbf{z}\right)$
for the parameters conditional on this allocation. Unfortunately, 
the computational burden is of order $\text{O}(J^n)$. This is even more frustrating 
when considering that the overwhelming majority of the posterior probabilities
$\omega\left(\mathbf{z}\right)$ will be close to zero for any sample.

\subsection{Exact posterior computation}\label{sec:xact}
In a somewhat paradoxical twist, we now proceed to show that, in some very special cases,
there exist exact derivations for the posterior distribution! This surprising phenomenon
only takes place for discrete distributions under a particular choice of the component
densities $f(x|\theta_i)$.
In essence, the $f(x|\theta_i)$'s must belong to the natural exponential families, i.e.
$$
f(x|\theta_i) = h(x)\exp\left\{ \theta_i\cdot R(x) - \Psi(\theta_i) \right\}\,,
$$
to allow for sufficient statistics
to be used. In this case, there exists a {\em conjugate prior} \citep{robert:2001} associated
with each $\theta$ in $f(x|\theta)$ as well as for the weights of the mixture. Let us consider the 
complete likelihood
\begin{align*}
\text{L}^c(\boldsymbol{\theta},\mathbf{p}|\bx,\bz) &= \prod_{i=1}^n p_{z_i}\,
\exp\left\{ \theta_{z_i}\cdot R(x_i) - \Psi(\theta_{z_i}) \right\} \\
&= \prod_{j=1}^J p_j^{n_j} \exp\left\{ \theta_j\cdot \sum_{z_i=j}
R(x_i) - n_j\Psi(\theta_j) \right\} \\
&= \prod_{j=1}^J p_j^{n_j} \exp\left\{ \theta_j\cdot S_j - n_j\Psi(\theta_j) \right\}\,,
\end{align*}
where $S_j=\sum_{z_i=j}R(x_i)$. It is easily seen that we remain in an exponential family since there exist sufficient
statistics with fixed dimension, $(n_1,\ldots,n_J,S_1,\ldots,S_J)$. Using a Dirichlet prior
$$
\pi(p_1,\ldots,p_J) = \frac{\Gamma(\alpha_1+\ldots+\alpha_J)}
                           {\Gamma(\alpha_1)\cdots\Gamma(\alpha_J)}
                                 p_1^{\alpha_1-1}\cdots p_J^{\alpha_J-1}
$$
on the vector of the weights $(p_1,\ldots,p_J)$ defined on the simplex of $\mathbb{R}^J$
and (independent) conjugate priors on the $\theta_j$'s,
$$
\pi(\theta_j) \propto \exp\left\{ \theta_j\cdot \tau_j - \delta_j\Psi(\theta_j) \right\} \,,
$$
the posterior associated with the complete likelihood $\text{L}^c(\boldsymbol{\theta},\mathbf{p}|\bx,\bz)$ is then of the same
family as the prior:
\begin{align*}
\pi(\boldsymbol{\theta},\mathbf{p}|\bx,\bz) &\propto \pi(\boldsymbol{\theta},\mathbf{p})\times \text{L}^c(\boldsymbol{\theta},\mathbf{p}|\bx,\bz)\\
&\propto \prod_{j=1}^J p_j^{\alpha_j-1}\,\exp\left\{ \theta_j\cdot \tau_j - \delta_j\Psi(\theta_j) \right\}\\
&\qquad\qquad \times p_j^{n_j} \exp\left\{ \theta_j\cdot S_j - n_j\Psi(\theta_j) \right\} \\
&=\prod_{j=1}^J p_j^{\alpha_j+n_j-1}\,\exp\left\{ \theta_j\cdot (\tau_j+S_j)
- (\delta_j+n_j)\Psi(\theta_j) \right\}\,;
\end{align*}
the parameters of the prior get transformed from $\alpha_j$ to $\alpha_j+n_j$,
from $\tau_j$ to $\tau_j+S_j$ and from $\delta_j$ to $\delta_j+n_j$.

If we now consider the observed likelihood (instead of the complete likelihood), it is the
sum of the complete likelihoods over all possible configurations of the partition space
of allocations, that is, a sum over $J^n$ terms,
$$
\sum_{\bz} \prod_{j=1}^J p_j^{n_j} \exp\left\{ \theta_j\cdot S_j - n_j\Psi(\theta_j) \right\}\,.
$$ 
The associated posterior is then, up to a constant,
\begin{align*}
\sum_{\bz} &\prod_{j=1}^J p_j^{n_j+\alpha_j-1} \exp\left\{ \theta_j\cdot (\tau_j+S_j)
        - (n_j+\delta_j)\Psi(\theta_j) \right\} \\
&= \sum_{\bz}\, \omega(\bz)\, \pi(\boldsymbol{\theta},\mathbf{p}|\bx,\bz)\,,
\end{align*}
where $\omega(\bz)$ is the normalising constant that is missing in
$$
\prod_{j=1}^J p_j^{n_j+\alpha_j-1} \exp\left\{ \theta_j\cdot (\tau_j+S_j)
        - (n_j+\delta_j)\Psi(\theta_j) \right\}\,.
$$
The weight $\omega(\mathbf{z})$ is therefore
$$
\omega(\bz)\propto \frac{\prod_{j=1}^J \Gamma(n_j+\alpha_j)}{\Gamma(\sum_{j=1}^J \{n_j+\alpha_j\})}\times
\prod_{j=1}^J K(\tau_j+S_j,n_j+\delta_j)\,,
$$
if $K(\tau,\delta)$ is the normalising constant of
$\exp\left\{ \theta_j\cdot \tau - \delta \Psi(\theta_j) \right\}$, i.e.
$$
K(\tau,\delta) = \int \exp\left\{ \theta_j\cdot \tau - \delta \Psi(\theta_j) \right\} \dd\theta_j\,.
$$

Unfortunately, except for very few cases, like Poisson and Multinomial mixtures, 
this sum does not simplify into a smaller number of terms because there exist no summary
statistics. From a Bayesian point of view, the complexity of the model is
therefore truly of magnitude $\text{O}(J^n)$.

We process here the cases of both the Poisson and Multinomial mixtures, noting that
the former case was previously exhibited by \cite{fearnhead:2005}.

\begin{example}
\begin{em}
Consider the case of a two component Poisson mixture, 
$$
x_1,\ldots,x_n\stackrel{\text{iid}}{\sim}
p\,\mathcal{P}(\lambda_1) + (1-p)\,\mathcal{P}(\lambda_2)\,,
$$
with a uniform prior on $p$ and exponential priors
$\mathcal{E}xp(\tau_1)$ and $\mathcal{E}xp(\tau_2)$ on $\lambda_1$ and $\lambda_2$, respectively.
For such a model, $S_j=\sum_{z_i=j} x_i$ and the normalising constant is then equal to
\begin{align*}
K(\tau,\delta) &= \int_{-\infty}^\infty \exp\left\{ \lambda_j \tau 
	- \delta \log(\lambda_j) \right\} \dd\lambda_j\\
   &= \int_0^\infty \lambda_j^{\tau-1}\,\exp \{-\delta\lambda_j\}\,\dd\lambda_j = \delta^{-\tau}\,\Gamma(\tau)\,.
\end{align*}
The corresponding posterior is (up to the overall normalisation of the weights)
\begin{align*}
\sum_{\bz}\, &\frac{\ds \prod_{j=1}^2 \Gamma(n_j+1)\Gamma(1+S_j)\big/(\tau_j+n_j)^{S_j+1}}{
	\Gamma(2+\sum_{j=1}^2 n_j)}\, \pi(\boldsymbol{\theta},\mathbf{p}|\bx,\bz) \\
&= \sum_{\bz}\, \frac{\ds \prod_{j=1}^2 n_j!\,S_j! \big/(\tau_j+n_j)^{S_j+1}}{
	(n+1)!}\, \pi(\boldsymbol{\theta},\mathbf{p}|\bx,\bz) \\
&\propto \sum_{\bz}\, \prod_{j=1}^2 \frac{n_j!\,S_j! }{(\tau_j+n_j)^{S_j+1}} \, 
	\pi(\boldsymbol{\theta},\mathbf{p}|\bx,\bz)\,.
\end{align*}
$\pi(\boldsymbol{\theta},\mathbf{p}|\bx,\bz)$ corresponds to a $\mathcal{B}(1+n_j,1+n-n_j)$ 
(Beta distribution) on $p_j$ and to a $\mathcal{G}(S_j+1,\tau_j+n_j)$ (Gamma distribution) 
on $\delta_j$, $(j=1,2)$. \\
An important feature of this example is that the above sum does not involve all of the $2^n$ terms,
simply because the individual terms factorise in $(n_1,n_2,S_1,S_2)$ that
act like local sufficient statistics. Since
$n_2=n-n_1$ and $S_2=\sum x_i - S_1$, the posterior only requires as many distinct terms as there
are distinct values of the pair $(n_1,S_1)$ in the completed sample. For instance, if the sample is
$(0,0,0,1,2,2,4)$, the distinct values of the pair $(n_1,S_1)$ are $(0,0),(1,0),(1,1),(1,2),(1,4),
(2,0),\allowbreak(2,1),\allowbreak(2,2),\allowbreak(2,3),
\allowbreak(2,4),(2,5),(2,6),\allowbreak\ldots,(6,5),\allowbreak(6,7), \allowbreak(6,8),\allowbreak(7,9)$. 
Hence there are $41$ distinct terms in the posterior, rather than $2^8=256$.\hfill$\blacktriangleleft$
\end{em}
\end{example}

Let $\mathbf{n}=\left(n_1,\ldots,n_J\right)$ and $\mathbf{S}=\left(S_1,\ldots,S_J\right)$.
The problem of computing the number (or cardinal) $\mu_n(\mathbf{n},\mathbf{S})$
of terms in the sum with an identical statistic $(\mathbf{n},\mathbf{S})$ has been tackled
by \cite{fearnhead:2005}, who proposes a recurrent formula to compute
$\mu_n(\mathbf{n},\mathbf{S})$ in an efficient book-keeping technique, as expressed below for a
$k$ component mixture:

\textit{If $\mathbf{e}_j$ denotes the vector of length $J$ made of zeros everywhere
except at component $j$ where it is equal to one, if 
$$
\mathbf{n}=(n_1,\ldots,n_J)\,,\quad\text{and}\quad \mathbf{n}-\mathbf{e}_j=(n_1,\ldots,n_j-1,\ldots,n_J)\,, 
$$
then
$$
\mu_1(\mathbf{e}_j,R\left(x_1\right)\mathbf{e}_j) = 1\,,\quad \forall j\in\{1,\ldots,J\}\,,\quad\text{and}\quad
\mu_n(\mathbf{n},\mathbf{S})=\sum_{j=1}^J \mu_{n-1}(\mathbf{n}
-\mathbf{e}_j,\mathbf{S}-R\left(x_n\right)\mathbf{e}_j)\,.
$$
}

\begin{example}
\begin{em}
Once the $\mu_n(\mathbf{n},\mathbf{S})$'s are all recursively computed, the posterior can be written as
$$
\sum_{(\mathbf{n},\mathbf{S})}\, \mu_n(\mathbf{n},\mathbf{S}) \prod_{j=1}^2 n_j!\,S_j! \big/(\tau_j+n_j)^{S_j+1}
\,\pi(\boldsymbol{\theta},\mathbf{p}|\bx,\mathbf{n},\mathbf{S})\,,
$$
up to a constant, and the sum only depends on the possible values of the ``sufficient" statistic $(\mathbf{n},\mathbf{S})$. \\
This closed form expression allows for a straightforward representation of the marginals. 
For instance, up to a constant, the marginal in $\lambda_1$ is given by
\begin{align*}
\sum_{\bz}\, &\prod_{j=1}^2 n_j!\,S_j! \big/(\tau_j+n_j)^{S_j+1}
\,(n_1+1)^{S_1+1}\lambda^{S_1}\,\exp\{-(n_1+1)\lambda_1\} / n_1!\\
&=\sum_{(\mathbf{n},\mathbf{S})}\mu_n(\mathbf{n},\mathbf{S}) \,\prod_{j=1}^2 n_j!\,S_j! \big/(\tau_j+n_j)^{S_j+1}
\\
&\qquad\times (n_1+\tau_1)^{S_1+1}\lambda_1^{S_1}\,\exp\{-(n_1+\tau_1)\lambda_1\} / n_1! \,.
\end{align*}
The marginal in $\lambda_2$ is
\begin{align*}
\sum_{(\mathbf{n},\mathbf{S})}\,&\mu_n(\mathbf{n},\mathbf{S}) \,\prod_{j=1}^2 n_j!\,S_j! \big/(\tau_j+n_j)^{S_j+1} \\
&\qquad(n_2+\tau_2)^{S_2+1}\lambda_2^{S_2}\,\exp\{-(n_2+\tau_2)\lambda_2\} / n_2!\,,
\end{align*}
again up to a constant.

Another interesting outcome of this closed form representation is that marginal
densities can also be computed in closed form. The marginal distribution
of $\bx$ is directly related to the unnormalised weights in
that
\begin{equation*}
m(\bx) = \sum_\bz \omega(\bz) = \sum_{(\mathbf{n},\mathbf{S})} \mu_n(\mathbf{n},\mathbf{S})\, \frac{\prod_{j=1}^2 n_j!\,S_j!
	\big/ (\tau_j+n_j)^{S_j+1}}{(n+1)!}
\end{equation*}
up to the product of factorials $1/x_1!\cdots x_n!$ (but this product is irrelevant in the
computation of the Bayes factor). \hfill$\blacktriangleleft$
\end{em}
\end{example}

Now, even with this considerable reduction in the complexity of the posterior distribution,
the number of terms in the posterior still explodes fast both with $n$ and with the
number of components $J$, as shown through a few simulated examples in Table \ref{tab:eXplose}.
The computational pressure also increases with the range of
the data, that is, for a given value of $(J,n)$, the number of values of the sufficient
statistics is much larger when the observations are larger, as shown for instance in the first 
three rows of Table \ref{tab:eXplose}: a simulated Poisson $\mathcal{P}(\lambda)$ sample of size $10$ is
mostly made of $0$'s when $\lambda=.1$ but mostly takes different values when $\lambda=10$.
The impact on the number of sufficient statistics can be easily assessed when $J=4$. (Note
that the simulated dataset corresponding to $(n,\lambda)=(10,.1)$ in Table \ref{tab:eXplose}
happens to correspond to a simulated sample made only of $0$'s, which explains the $n+1=11$ values of
the sufficient statistic $(n_1,S_1)=(n_1,0)$ when $J=2$.)

\begin{table}\begin{center}
\begin{tabular}{l|ccc}
$(n,\lambda)$   & $J=2$ & $J=3$ & $J=4$\\
\hline
$(10,.1)$       &  11   &   66    &  286 \\
$(10,1)$        &  52   &  885    & 8160 \\
$(10,10)$       & 166   & 7077    & 120,908\\
$(20,.1)$       &  57   &  231    & 1771 \\
$(20,1)$        & 260   & 20,607  & 566,512\\
$(20,10)$       & 565   & 100,713 &  --- \\
$(30,.1)$       &  87   &  4060   & 81,000\\
$(30,1)$        & 520   & 82,758  &  --- \\
$(30,10)$       & 1413  & 637,020 &  --- \\
%$(40,.1)$       & 216   & 13,986  &  --- \\
%$(40,1)$        & 789   & 271,296 &  --- \\
%$(40,10)$       & 2627  &  ---    &  --- \\
\hline
\end{tabular}
\caption{\label{tab:eXplose}
Number of pairs $(\mathbf{n},\mathbf{S})$ for simulated datasets
from a Poisson $\mathcal{P}(\lambda)$ and different
numbers of components. {\em (Missing terms are due
to excessive computational or storage requirements.)}
}
\end{center}\end{table}

\begin{example}
\begin{em}
If we have $n$ observations $\bn_i=(n_{i1},\ldots,n_{im})$ from the 
Multinomial mixture
$$
\bn_i\sim
p\mathcal{M}_m(d_i;q_{11},\ldots,q_{1m})+(1-p)\mathcal{M}_m(d_i;q_{21},\ldots,q_{2m})\,
$$
where $n_{i1}+\cdots+n_{im}=d_i$ and $q_{11}+\cdots+q_{1m}=q_{21}+\cdots+q_{2m}=1$, 
the conjugate priors on the $q_{jv}$'s are Dirichlet distributions, $(j=1,2)$
$$
(q_{j1},\ldots,q_{jm}) \sim \mathcal{D}(\alpha_{j1},\ldots,\alpha_{jm})\,,
$$
and we use once again the uniform prior on $p$. (A default choice for the $\alpha_{jv}$'s
is $\alpha_{jv}=1/2$.) Note that the $d_j$'s may differ from observation to observation, since
they are irrelevant for the posterior distribution: given a partition $\bz$ of the sample,
the complete posterior is indeed 
$$
p^{n_1} (1-p)^{n_2}\,\prod_{j=1}^2\prod_{z_i=j}q_{j1}^{n_{i1}}\cdots q_{jm}^{n_{im}}\,
\times \prod_{j=1}^2\prod_{v=1}^m q_{jv}^{-1/2},
$$
(where $n_j$ is the number of observations allocated to component $j$)
up to a normalising constant that does not depend on $\bz$. \hfill$\blacktriangleleft$
\end{em}
\end{example}

More generally, considering a Multinomial mixture with $m$ components,
$$
\bn_i\sim \sum_{j=1}^J 
p_j\mathcal{M}_m(d_i;q_{j1},\ldots,q_{jm})\,,
$$
the complete posterior is also directly available, as
$$
\prod_{j=1}^J p_j^{n_j}\times
\prod_{j=1}^J\prod_{z_i=j}q_{j1}^{n_{i1}}\cdots q_{jm}^{n_{im}}\,
\times \prod_{j=1}^J\prod_{v=1}^m q_{jv}^{-1/2},
$$
once more up to a normalising constant.

Since the corresponding normalising constant of the Dirichlet distribution is
$$
\frac{\prod_{v=1}^m \Gamma(\alpha_{jv}) }{\Gamma(\alpha_{j1}+\cdots+\alpha_{jm})}\,,
$$
the overall weight of a given partition $\bz$ is
\begin{equation}\label{eq:normaM}
n_1!n_2!\frac{\prod_{v=1}^m \Gamma(\alpha_{1v}+S_{1v})}{\Gamma(\alpha_{11}
+\cdots+\alpha_{1m}+S_{1\cdot})}\times\frac{\prod_{v=1}^m 
\Gamma(\alpha_{2v}+S_{2v})}{\Gamma(\alpha_{21} +\cdots+\alpha_{2m}+S_{2\cdot})}\,
\end{equation}
where $S_{ji}$ is the sum of the $n_{ji}$'s for the observations $i$
allocated to component $j$ and 
$$
S_{ji}=\sum_{z_i=j} n_{ji}\quad\text{and}\quad S_{j\cdot}=\sum_i S_{ji}\,.
$$ 

Given that the posterior distribution only depends on those ``sufficient" statistics 
$S_{i j}$ and $n_i$, the same factorisation as in the Poisson case applies, namely we simply 
need to count the number of occurrences of a particular local sufficient statistic 
$(n_{1},S_{11}, \ldots,S_{Jm})$ and then sum over all values of this sufficient
statistic. The book-keeping algorithm of \cite{fearnhead:2005} applies.
Note however that the number of different terms in the closed form expression is growing
extremely fast with the number of observations, with the number of components and with
the number $k$ of modalities.

\begin{example}\label{exa:stoufftob}
\begin{em}
In the case of the latent class model, consider the simplest case of two variables with two
modalities each, so observations are products of Bernoulli's,
$$
\bx\sim p\mathcal{B}(q_{11})\mathcal{B}(q_{12})+(1-p)\mathcal{B}(q_{21})\mathcal{B}(q_{22})\,.
$$
We note that the corresponding statistical model is not identifiable
beyond the usual label switching issue detailled in Section \ref{sub:ifaible}.
Indeed, there are only two dichotomous variables, four possible realizations for the $\bx$'s,
and five unknown parameters. We however take advantage of this artificial model 
to highlight the implementation of the above exact algorithm, which can then easily uncover
the unidentifiability features of the posterior distribution.

The complete posterior distribution is the sum over all partitions of the terms
$$
p^{n_1} (1-p)^{n_2}\,\prod_{j=1}^2\,\prod_{v=1}^2\,q_{jv}^{s_{jv}}(1-q_{jv})^{n_j-s_{jv}}
\times \prod_{j=1}^2\prod_{v=1}^2 q_{jv}^{-1/2}
$$
where $s_{jv}=\sum_{z_i=j}x_{iv}$, the sufficient statistic is thus $(n_1,s_{11},s_{12},s_{21},s_{22})$,
of order $\text{O}(n^5)$. Using the benchmark data of \cite{stouffer:toby:1951}, made of 216 sample points
involving four binary variables related with a sociological questionnaire, we
restricted ourselves to both first variables and 50 %the 200 first 
observations picked at random. A recursive algorithm that
eliminated replicates gives the results that (a) there are $5,928$ different values for the sufficient statistic
and (b) the most common occurrence is the middle partition $(26,6,11,5,10)$,
with $7.16\times10^{12}$ replicas (out of $1.12\times 10^{15}$ total partitions). 
The posterior weight of a given partition is
\begin{align*}
&\frac{\Gamma(n_1+1)\Gamma(n-n_1+1)}{\Gamma(n+2)}\,
\prod_{j=1}^2\,\prod_{v=1}^2 \frac{\Gamma(s_{jv}+1/2)\Gamma(n_j-s_{jv}+1/2)}{\Gamma(n_j+1)}\\
&\quad=\prod_{j=1}^2\,\prod_{v=1}^2 \Gamma(s_{jv}+1/2)\Gamma(n_j-s_{jv}+1/2) \bigg/ 
n_1!\,(n-n_1)!\,(n+1)!\,,
\end{align*}
multiplied by the number of occurrences. In this case, it is therefore possible to find 
exactly the most likely partitions, namely the one with
$n_1=11$ and $n_2=39$, $s_{11}=11$, $s_{12}=8$, $s_{21}=0$, $s_{22}=17$, and the symmetric
one, which both only occur once and which have a joint posterior probability of $0.018$.
It is also possible to eliminate all the partitions with very low probabilities in this example. \hfill$\blacktriangleleft$
\end{em}
\end{example} 

%%%%%%%%%%%%%%%%%%%%%%%%%%%%%%%%%%%%%%%%%%%%%%%%%%%%
\section{Mixture inference}\label{sec:conundrum}
%%%%%%%%%%%%%%%%%%%%%%%%%%%%%%%%%%%%%%%%%%%%%%%%%%%%

Once again, the apparent simplicity of the mixture density should not be
taken at face value for inferential purposes; since, for a sample
of arbitrary size $n$ from a mixture distribution \eqref{eq:mix}, 
there always is a non-zero probability $(1-p_j)^n$ that the $j$th subsample is empty,
the likelihood includes terms that do not bring any
information about the parameters of the $i$-th component. 

\subsection{Nonidentifiability, hence label switching}\label{sub:ifaible}
A mixture model \eqref{eq:mix} is {\em senso stricto} never identifiable 
since it is invariant under permutations of the indices of the components. 
Indeed, unless we introduce some restriction on the range of the
$\theta_i$'s, we cannot distinguish component number $1$ (i.e., $\theta_1$) from component 
number $2$ (i.e., $\theta_2$) in the likelihood, because they are exchangeable. 
This apparently benign feature has consequences on both Bayesian inference and
computational implementation. First, exchangeability implies that in a $J$ component mixture, 
the number of modes is of order $\mathrm{O}(J!)$. The highly multimodal posterior surface is 
therefore difficult to explore via standard Markov chain Monte Carlo techniques.
Second, if an exchangeable prior is used on $\boldsymbol{\theta}=\left(\theta_1,\ldots,\theta_J\right)$,
all the marginals of the $\theta_j$'s are identical. Other and more severe sources of unidentifiability 
could occur as in Example \ref{exa:stoufftob}.

\begin{figure}
\begin{center}\includegraphics[width=180pt]{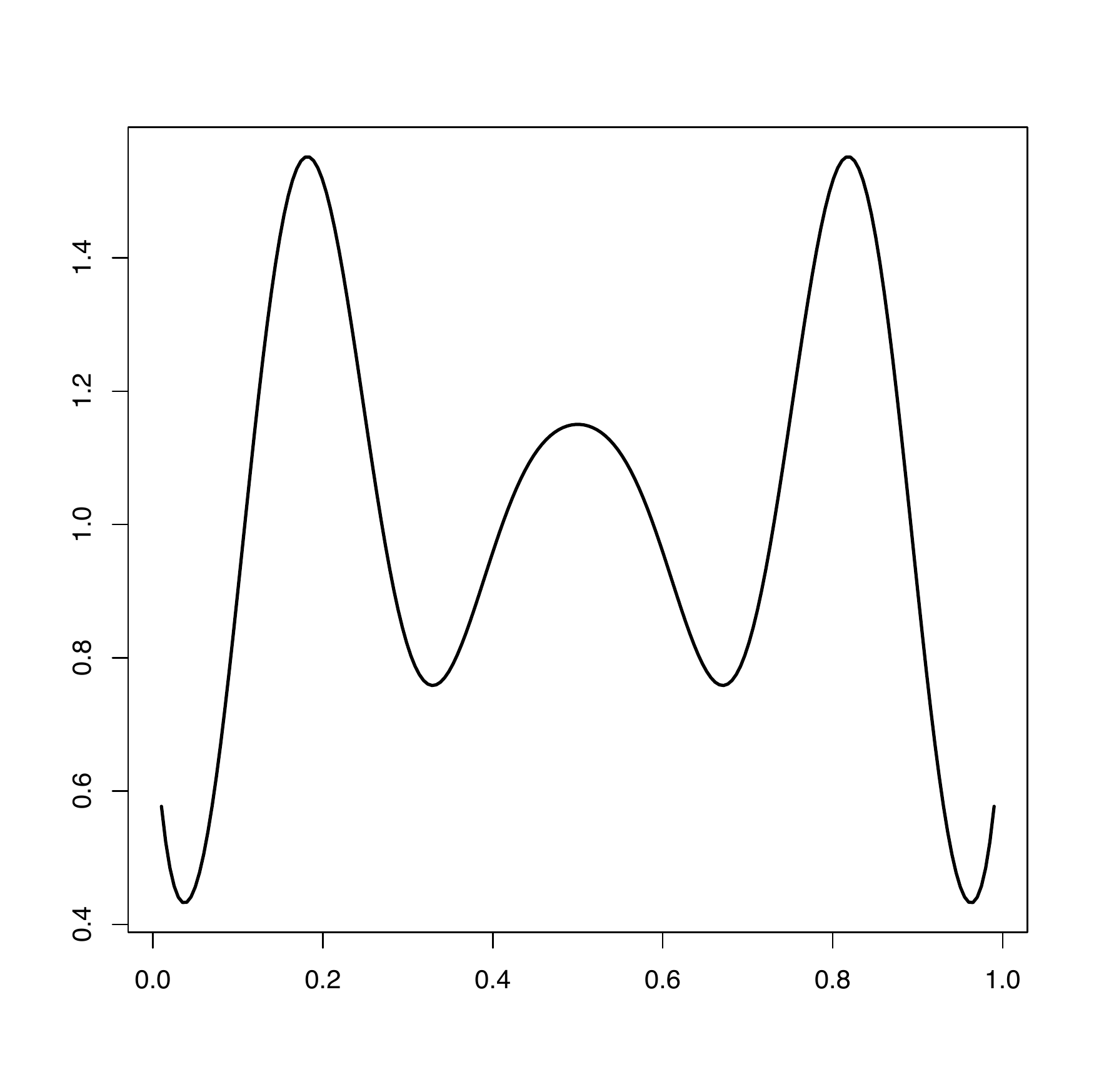}\end{center}
\caption{\label{fig:XaXul}
Exact posterior distribution of $p$ for a sample of $50$ observations from the dataset
of \cite{stouffer:toby:1951} and $a=b=1$.
}
\end{figure}

\begin{figure}
\begin{center}\includegraphics[width=180pt]{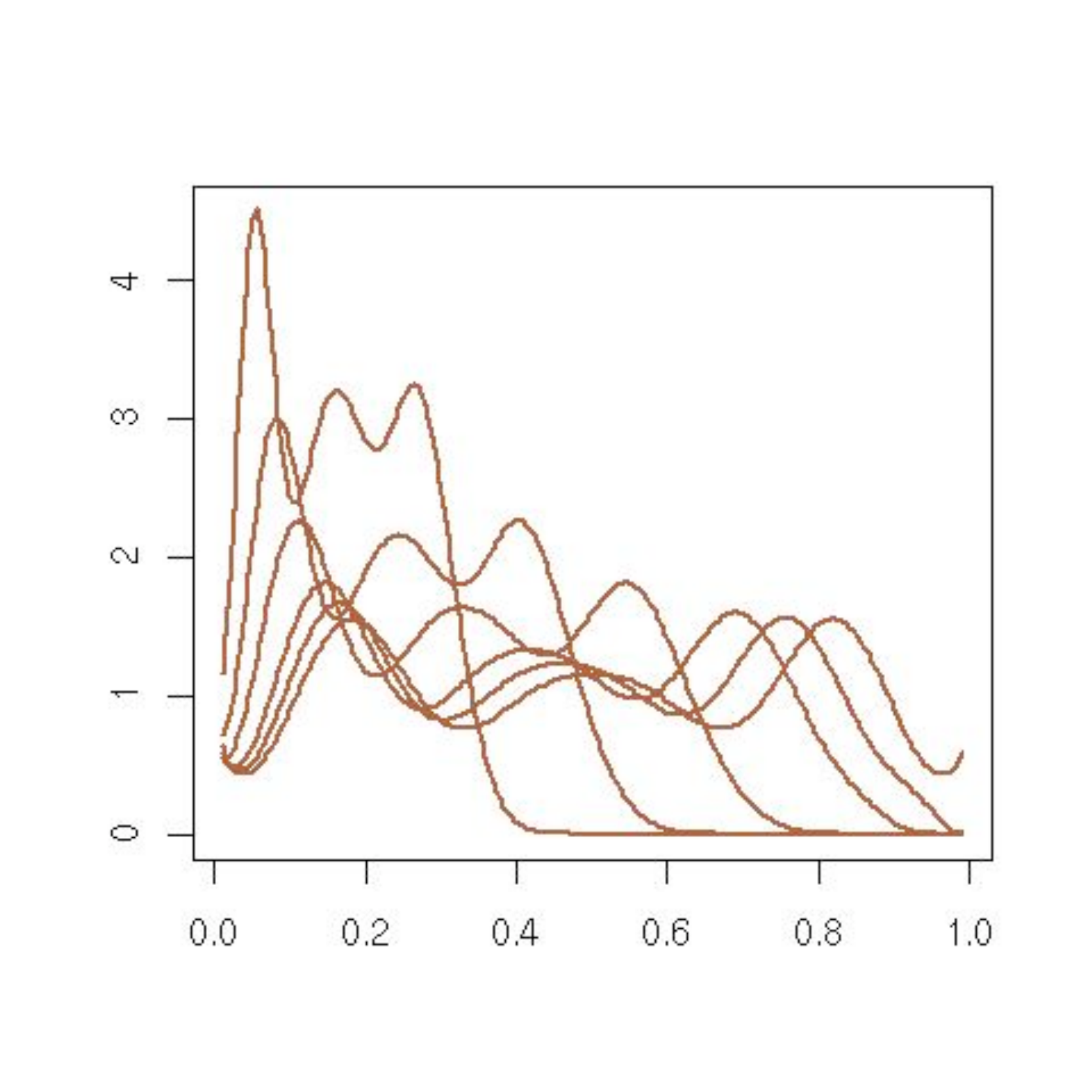}\end{center}
\caption{\label{fig:noiden2}
Exact posterior distributions of $p$ for a sample of 50 observations from the dataset of \cite{stouffer:toby:1951}
under Beta $\mathcal{B}(a,b)$ priors when $a=.01,.05,.1,.05,1$ and $b=100,50,20,10,5,1$.}
\end{figure}

\begin{example} \begin{em} {\bf (Example \ref{exa:stoufftob} continued)}
If we continue our assessment of the latent class model, with two variables with two
modalities each, based on subset of data extracted from \cite{stouffer:toby:1951}, under a Beta, $\mathcal{B}(a,b)$, prior distribution
on $p$ the posterior distribution is the weighted sum of Beta $\mathcal{B}(n_1+a,n-n_1+b)$
distributions, with weights 
$$
\mu_n(\mathbf{n},\mathbf{s})\,\prod_{j=1}^2\,\prod_{v=1}^2 \Gamma(s_{jv}+1/2)
\Gamma(n_j-s_{jv}+1/2) \bigg/ n_1!\,(n-n_1)!\,(n+1)!\,,
$$
where $\mu_n(\mathbf{n},\mathbf{s})$ denotes the number of occurrences of the sufficient
statistic.  Figure \ref{fig:XaXul} provides the posterior distribution for a subsample of 
the dataset of \cite{stouffer:toby:1951} and $a=b=1$.
Since $p$ is not identifiable, the impact of the prior distribution is stronger than in an identifying setting:
using a Beta $\mathcal{B}(a,b)$ prior on $p$ thus produces a posterior [distribution] that reflects as much the influence
of $(a,b)$ as the information contained in the data. While a $\mathcal{B}(1,1)$ prior, as in Figure \ref{fig:XaXul}, 
leads to a perfectly symmetric posterior with three modes, using an assymetric prior with $a\ll b$ strongly modifies the range of the posterior,
as illustrated by Figure \ref{fig:noiden2}. \hfill$\blacktriangleleft$
\end{em}
\end{example}

Identifiability problems resulting from the exchangeability issue are called ``label switching" in
that the output of a properly converging MCMC algorithm should produce no information
about the component labels (a feature which, incidentally, provides a fast assessment of the performance of MCMC solutions,
as proposed in \citealp{Celeux:Hurn:Robert:2000}).
A na\"\i ve answer to the problem proposed in the early literature is to impose an {\em identifiability 
constraint} on the parameters, for instance by ordering the means (or the variances or the weights) 
in a normal mixture. From a Bayesian point of view, this amounts to truncating
the original prior distribution, going from $\pi\left(\boldsymbol{\theta},\mathbf{p}\right)$ to
$$
\pi\left(\boldsymbol{\theta},\mathbf{p}\right)\,\mathbb{I}_{\mu_1\le\ldots\le\mu_J}\,.
$$

While this device may seem innocuous (because indeed the sampling distribution is
the same with or without this constraint on the parameter space), it is not without
consequences on the resulting inference. This can be seen directly on the posterior
surface: if the parameter space is reduced to its constrained
part, there is no agreement between the above notation and the topology of this surface.
Therefore, rather than selecting a single posterior mode and its neighbourhood, the constrained
parameter space will most likely include parts of several modal regions. Thus, the resulting posterior
mean may well end up in a very low probability region and be unrepresentative of the estimated distribution.

Note that, once an MCMC sample has been simulated from an unconstrained posterior distribution, 
any ordering constraint can be imposed on this sample, that is, after the simulations have been 
completed, for estimation purposes as stressed by \cite{Stephens:1997}. 
Therefore, the simulation (if not the estimation) hindrance created
by the constraint can be completely bypassed.

%The strong effect of constraining the parameter space is illustrated by Table 
%\ref{tab:nonsense} of \cite{Celeux:Hurn:Robert:2000}:
%Depending on which parameter order is selected, the estimators vary widely and produce densities that
%are unrelated with the truth \citep[see also][]{marin:mengersen:robert:2005}.
%\begin{table}
%\begin{center}
%{\small
%\begin{tabular}{l|ccc ccc ccc}
%\hline
%$\hbox{order}$&$p_1$ &$p_2$ &$p_3$ &$\theta_1$ &$\theta_2$ &$\theta_3$ &$\sigma_1$ &$\sigma_2$ &$\sigma_3$\cr
%\hline
%$p$ &0.231 &0.311 &0.458 &0.321 &-0.55 &2.28 &0.41 &0.471 &0.303\\
%$\theta$ &0.297 &0.246 &0.457 &-1.1 &0.83 &2.33 &0.357 &0.543 &0.284\\
%$\sigma$ &0.375 &0.331 &0.294 &1.59 &0.083 &0.379 &0.266 &0.34 &0.579\\
%$\hbox{true}$ &0.22 &0.43 &0.35 &1.1 &2.4 &-0.95 &0.3 &0.2 &0.5\\
%\hline
%\end{tabular}
%}\end{center}
%\caption{\label{tab:nonsense}
%Estimates of the parameters of a three component normal mixture 
%for a simulated sample of $500$ points, obtained by re-ordering 
%an MCMC sample according to one of three constraints, $p:\,p_1<p_2<p_3$,
%$\mu:\,\mu_1<\mu_2<\mu_3$, or $\sigma:\,\sigma_1<\sigma_2<\sigma_3$.
%({\em Source:} \citealp{Celeux:Hurn:Robert:2000})}
%\end{table}

Once an MCMC sample has been simulated from an unconstrained posterior distribution,
a natural solution is to identify one of the $J!$ modal
regions of the posterior distribution and to operate the relabelling in terms of proximity to this
region, as in \cite{marin:mengersen:robert:2005}.
Similar approaches based on clustering algorithms for the parameter sample
are proposed in \cite{Stephens:1997} and \cite{Celeux:Hurn:Robert:2000}, and
they achieve some measure of success on the examples for which they have been tested. 

\subsection{Restrictions on priors}\label{sub:mix.prior}
From a Bayesian point of view, the fact that few or no observation in the sample is (may be) generated from 
a given component has a direct and important drawback: this prohibits the use of independent improper priors,
$$\pi\left(\boldsymbol{\theta}\right)=\prod_{j=1}^J\pi(\theta_j)\,,$$
since, if 
$$\int\pi(\theta_j)\hbox{d}\theta_j=\infty$$
then for any sample size $n$ and any sample $\mathbf{x}$, 
$$\int\pi(\boldsymbol{\theta},\mathbf{p}|\mathbf{x}) 
\hbox{d}\boldsymbol{\theta}\hbox{d}\mathbf{p}=\infty\,.$$
%because, among the $k^n$ terms in the expansion of $\pi(\underline{\theta},\underline{p}|\underline{x})$,
%there are $(k-1)^n$ with {\em no} observation allocated to the $i$-th component
%and thus a conditional posterior $\pi(\theta_i|\underline{x},\underline{z})$ equal to the
%prior $\pi_i(\theta_i)$.

The ban on using improper priors can be considered by some as being of little importance, 
since proper priors with large variances could be used instead.
However, since mixtures are ill-posed problems, this difficulty with improper priors is more
of an issue, given that the influence of a particular proper prior, no matter how large
its variance, cannot be truly assessed.

There exists, nonetheless, a possibility of using improper priors in this setting, as demonstrated
for instance by \cite{Mengersen:Robert:1996}, by ad\-ding some degree of dependence between the
component parameters. In fact, a Bayesian perspective makes it quite easy to argue 
{\em against} independence in mixture models, since the components are only properly defined 
in terms of one another. For the very reason that 
exchangeable priors lead to identical marginal posteriors on all components, the relevant priors 
must contain some degree of information that components are {\em different} and those priors must
be explicit about this difference.

The proposal of \cite{Mengersen:Robert:1996}, also described in \cite{marin:mengersen:robert:2005}, 
is to introduce first a common reference, namely a scale, location, or location-scale
parameter $(\mu,\tau)$, and then to define the original parameters in terms of {\em departure} 
from those references. Under some conditions on the reparameterisation, 
expressed in \cite{Robert:Titterington:1998}, this representation allows 
for the use of an improper prior on the reference parameter $(\mu,\tau)$. 
See \cite{Wasserman:2000,Perez:Berger:2002,Moreno:Liseo:2003} for different 
approaches to the use of default or non-informative priors in the setting of mixtures.

%Reparametrisations have been developed for Gaussian mixtures \citep{Roeder:Wasserman:1997},
%and also for exponential \citep{Gruet:Philippe:Robert:1999} 
%and Poisson mixtures \citep{Robert:Titterington:1998}.
%However, these alternative representations do require the artificial identifiability 
%restrictions criticized above and less directly interpretable.

%%%%%%%%%%%%%%%%%%%%%%%%%%%%%%%%%%%%%%%%
\section{Inference for mixtures with a known number of components}
\label{sec:MCMC}
%%%%%%%%%%%%%%%%%%%%%%%%%%%%%%%%%%%%%%%%

In this section, we describe different Monte Carlo algorithms
that are customarily used for the approximation of posterior distributions in mixture settings
when the number of components $J$ is known. We start in Section \ref{sub:ReOr} with a proposed solution to the label-switching problem
and then discuss in the following sections Gibbs sampling and Metropolis-Hastings algorithms,
acknowledging that a diversity of other algorithms exist (tempering, population Monte Carlo...), see \cite{robert:casella:2004}.

\subsection{Reordering}\label{sub:ReOr}
Section \ref{sub:ifaible} discussed the drawbacks of imposing identifiability ordering constraints
on the parameter space for estimation performances and there are similar drawbacks on the
computational side, since those constraints decrease the explorative abilities of a sampler
and, in the most extreme cases, may even prevent the sampler from converging 
\citep[see][]{Celeux:Hurn:Robert:2000}. We thus consider samplers that evolve in an unconstrained parameter 
space, with the specific feature that the posterior surface has a number of modes that is a multiple of $J!$. 
Assuming that this surface is properly visited by the sampler (and this is not a trivial assumption), the derivation of point 
estimates of the parameters of (\ref{eq:mix}) follows from an \textit{ex-post} reordering
proposed by \cite{marin:mengersen:robert:2005} which we describe below.

Given a simulated sample of size $M$, a starting value for a point estimate is the na{\"\i}ve
approximation to the Maximum a Posteriori (MAP) estimator, that is the value in the
sequence $(\boldsymbol{\theta},\mathbf{p})^{(l)}$ that maximises the posterior,
$$
l^*=\arg\max_{l=1,\ldots,M} \pi((\boldsymbol{\theta},\mathbf{p})^{(l)}|\mathbf{x})
$$
Once an approximated MAP is computed, it is then possible to reorder all terms in the sequence
$(\boldsymbol{\theta},\mathbf{p})^{(l)}$ by selecting the reordering that is the closest to the approximate
MAP estimator for a specific distance in the parameter space. This solution
bypasses the identifiability problem without requiring a preliminary and most likely
unnatural ordering with respect to one of the parameters (mean, weight, variance) of the model.
Then, after the reordering step, an estimation of $\theta_j$ is given by 
$$
\sum_{l=1}^M(\theta_j)^{(l)}\big/M\,.
$$

\subsection{Data augmentation and Gibbs sampling approximations}

The Gibbs sampler is the most commonly used approach in Bayesian mixture estimation
\citep{Diebolt:Robert:1990b,Diebolt:Robert:1994,Lavine:West:1992,Verdinelli:Wasserman:1992, Escobar:West:1995}
because it takes advantage of the missing data structure of the $z_i$'s uncovered 
in Section \ref{sub:mix.miss}. 

The Gibbs sampler for mixture models (\ref{eq:mix}) \citep{Diebolt:Robert:1994} is based on 
the successive simulation of $\mathbf{z}$, 
$\mathbf{p}$ and $\boldsymbol{\theta}$ conditional on one another and on the data,
using the full conditional distributions derived from the conjugate structure of the
complete model. (Note that $\mathbf{p}$ only depends on the missing data $\bz$.)

\medskip
{\flushright{\bf Gibbs sampling for mixture models}}
%\colorbox{LightGrey}{\makebox[0.95\textwidth][c]{\parbox{0.95\textwidth}{
{\sf
\begin{enumerate}
\item[{\sf 0.}]  {\sffamily\bfseries Initialization:} 
		 choose $\mathbf{p}^{(0)}$ and $\boldsymbol{\theta}^{(0)}$ arbitrarily
\item[{\sf 1.}]  {\sffamily\bfseries Step t.} For $t=1,\ldots$
\begin{enumerate}
\item[{\sf 1.1}] Generate $z_i^{(t)}$ ($i=1,\ldots,n$) from ($j=1,\ldots,J$)
                 \begin{center}
                 $\mathbb{P}\left(z_i^{(t)}=j|p_j^{(t-1)},\theta_j^{(t-1)},x_i\right)\propto 
                 p_j^{(t-1)}f\left(x_i|\theta_j^{(t-1)}\right)$
                 \end{center}
\item[{\sf 1.2}] Generate $\mathbf{p}^{(t)}$ from $\pi(\mathbf{p}|\mathbf{z}^{(t)})$ 
\item[{\sf 1.3}] Generate $\boldsymbol{\theta}^{(t)}$ from $\pi(\boldsymbol{\theta}|\mathbf{z}^{(t)},\mathbf{x})$.
\end{enumerate}
\end{enumerate}
}
\medskip

As always with mixtures, the convergence of this MCMC algorithm is not as easy to assess as it seems at first sight. In fact, 
while the chain is uniformly geometrically ergodic from a theoretical point of view, the severe 
augmentation in the dimension of the chain brought by the completion stage may induce strong convergence problems.
The very nature of Gibbs sampling may lead to ``trapping sta\-tes", that is, concentrated local modes that
require an enormous number of iterations to escape from. For example, components with a small number of
allocated observations and very small variance become so tightly concentrated that there is very little 
probability of moving observations in or out of those components, as shown in \cite{marin:mengersen:robert:2005}. 
As discussed in Section \ref{sub:comics}, \cite{Celeux:Hurn:Robert:2000} show that most MCMC 
samplers for mixtures, including the Gibbs sampler, fail to reproduce the permutation
invariance of the posterior distribution, that is, that they do not visit the $J!$ replications of a given mode.

\begin{figure}
\begin{center}\includegraphics[width=240pt]{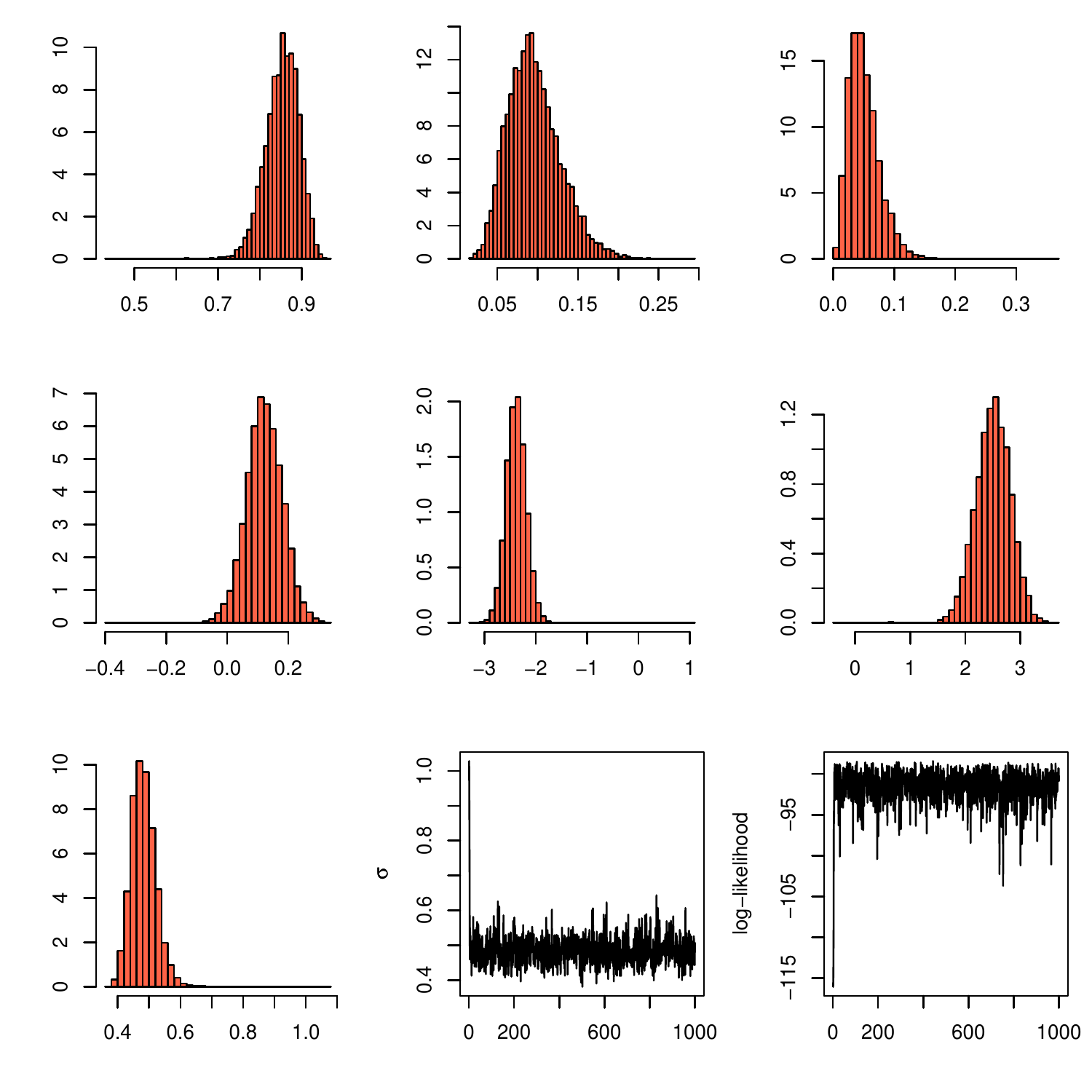}\end{center}
\caption{\label{fig:galax}
From the left to the right, histograms of the parameters $(p_1,p_2,p_3,\mu_1,\mu_2,\mu_3,\sigma)$ 
of a normal mixture with $k=3$ components based on $10^4$ iterations of the Gibbs sampler and
the galaxy dataset, evolution of the $\sigma$ and of the log-likelihood.
}
\end{figure}

\begin{example}
\begin{em}
\label{ex:galax} Consider a mixture of normal distributions with common variance $\sigma^2$ and unknown means and weights
$$
\sum_{j=1}^J p_j\,\mathcal{N}(\mu_j,\sigma^2)\,.
$$
This model is a particular case of model (\ref{eq:mix}) and is not identifiable.
Using conjugate exchangeable priors 
$$
\mathbf{p}\sim\mathcal{D}(1,\ldots,1)\,,\quad
\mu_j\sim\mathcal{N}(0,10\sigma^2)\,,\quad
\sigma^{-2}\sim\mathcal{E}xp(1/2)\,,
$$
it is straightforward to implement the above Gibbs sampler: 
\begin{itemize}
\item the weight vector $\mathbf{p}$ is simulated as the Dirichlet variable
$$
\mathcal{D}(1+n_1,\ldots,1+n_J)\,;
$$

\item the inverse variance as the Gamma variable
$$
\mathcal{G}\left\{(n+2)/2,(1/2)\left[1+\sum_{j=1}^J \left(
\frac{0.1 n_j \bar{x}_j^2}{n_j+0.1} +s_j^2\right) \right]\right\}\,;
$$

\item and, conditionally on $\sigma$, the means $\mu_j$ are simulated as the Gaussian variable
$$
\mathcal{N}(n_j \bar{x}_j/(n_j+0.1),\sigma^2/(n_j+0.1))\,;
$$
\end{itemize}
where $n_j=\sum_{z_i=j}$, $\bar x_j=\sum_{z_i=j}x_i$ and
$s_j^2=\sum_{z_i=j}(x_i-\bar x_j)^2/n_j$. \\
Note that this choice of implementation allows for the block simulation of the means-variance group, rather
than the more standard simulation of the means conditional on the variance and of the variance conditional
on the means \citep[as in][]{Diebolt:Robert:1994}. \hfill$\blacktriangleleft$

Consider the benchmark dataset of the galaxy radial speeds
described for instance in \cite{Roeder:Wasserman:1997}. 
The output of the Gibbs sampler is summarised on Figure
\ref{fig:galax} in the case of $J=3$ components. As is obvious from the comparison of the three first histograms
(and of the three following ones), label switching does not occur with this sampler: the three components remain
isolated during the simulation process. \hfill$\blacktriangleleft$
\end{em}
\end{example}

Note that \cite{Geweke:2007} (among others) dispute the relevance of asking for proper mixing
over the $k!$ modes, arguing that on the contrary the fact that the Gibbs sampler sticks to a single
mode allows for an easier inference. We obviously disagree with this perspective: first, from an algorithmic
point of view, given the unconstrained posterior distribution as the target, a sampler that fails to explore all 
modes clearly fails to converge. Second, the idea that being restricted to a single mode provides a
proper representation of the posterior is na\"{i}vely based on an intuition derived from mixtures with
few components. As the number of components increases, modes on the posterior surface get inextricably mixed and a standard MCMC
chain cannot be garanteed to remain within a single modal region. Furthermore, it is impossible to check in practice whether not
this is the case.

In his defence of ``simple" MCMC strategies supplemented with postprocessing steps, \cite{Geweke:2007} states that
\begin{quotation}\begin{em}
[Celeux et al.'s (2000)] argument is persuasive only to the extent that there are mixing problems beyond
those arising from permutation invariance of the posterior distribution. Celeux
et al.~(2000) does not make this argument, indeed stating ``The main defect of the
Gibbs sampler from our perspective is the ultimate attraction of the local modes''
(p. 959). That article produces no evidence of additional mixing problems in its 
examples, and we are not aware of such examples in the related literature. Indeed,
the simplicity of the posterior distributions conditional on state assignments in most
mixture models leads one to expect no irregularities of this kind.
\end{em}\end{quotation}
There are however clear irregularities in the convergence behaviour of Gibbs and Metropolis--Hastings
algorithms as exhibited in \cite{marin:mengersen:robert:2005} and \cite{marin:robert:2007} (Figure 6.4)
for an identifiable two-component normal mixture with both means unknown.
In examples as such as those, there exist secondary modes that may have much lower posterior values than
the modes of interest but that are nonetheless too attractive for the Gibbs sampler to visit other modes. In such
cases, the posterior inference derived from the MCMC output is plainly incoherent. (See also 
\cite{iacobucci:marin:robert:2008} for another illustration of a multimodal posterior distribution in
an identifiable mixture setting.) 

However, as shown by the example below, for identifiable mixture models, there is no label switching to expect
and the Gibbs sampler may work quite well. While there is no foolproof approach to check MCMC convergence \citep{robert:casella:2004}, we recommend using the
visited likelihoods to detect lack of mixing in the algorithms. This does not detect the label switching difficulties (but individual histograms do) but rather the possible
trapping of a secondary mode or simply the slow exploration of the posterior surface. This is particularly helpful when implementing
multiple runs in parallel.

\begin{figure}[ht!] 
\centerline{\includegraphics[width=12cm, height=4.5cm]{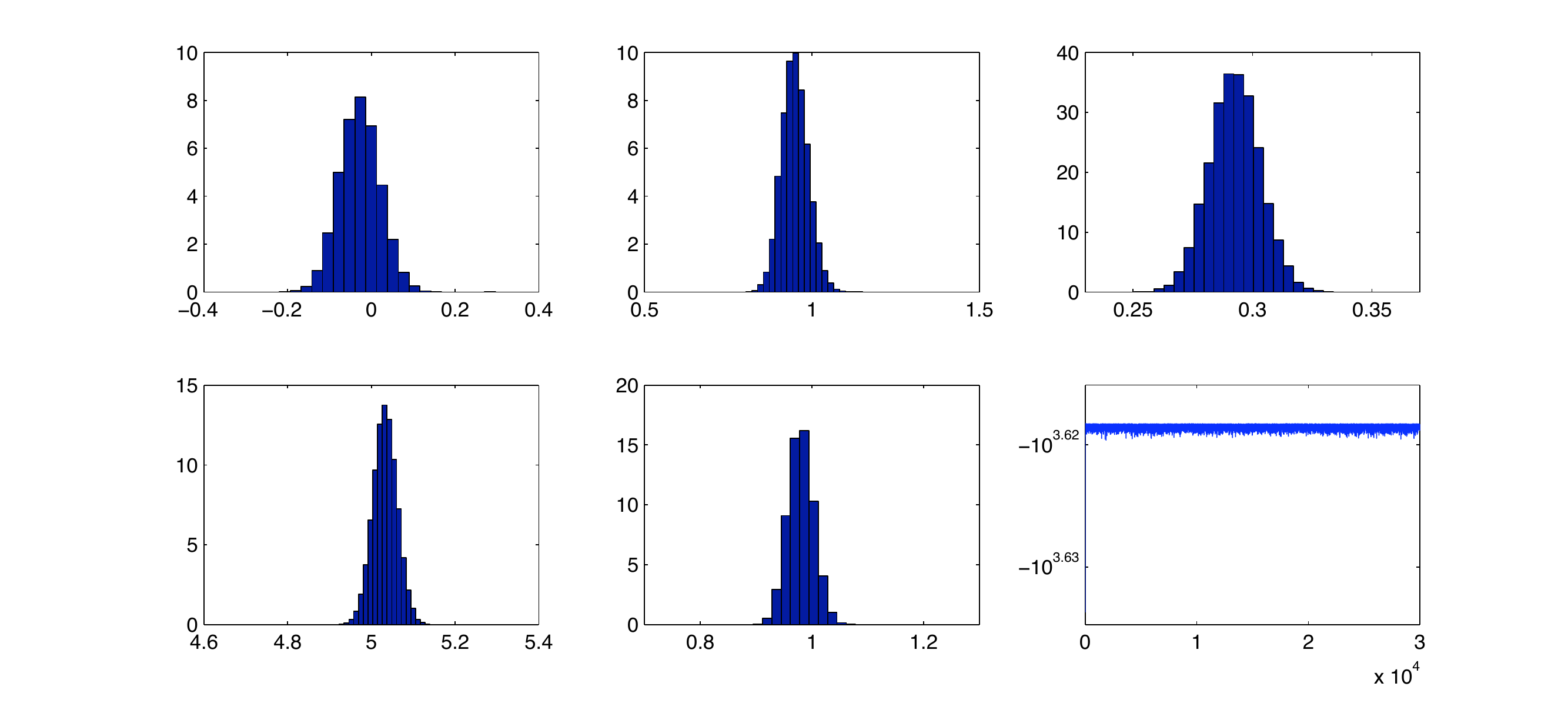}}
\caption{Histograms of the parameters, $\mu_1,\sigma_1, p_1,\mu_2,\sigma_2$, and evolution of the (observed) log-likelihood along
$30,000$ iterations of the Gibbs sampler and a sample of $2,000$ observations.}
\label{Gibbs_SimData}
\end{figure}

\begin{example}
\begin{em}
{\bf (Example \ref{ex:aero} continued)}
Consider the case of a mixture of Student's $t$ distributions with \textbf{known and different} numbers of degrees of
freedom
$$ 
\sum_{j=1}^J p_j t_{\nu_j}(\mu_j,\sigma_j^2)\,. 
$$
This mixture model is not a particular case of model (\ref{eq:mix}) and is identifiable. Moreover, since the noncentral $t$ distribution
$t_{\nu}(\mu,\sigma^2)$ can be interpreted  as a continuous mixture of normal distributions with a
common mean and with variances distributed as scaled inverse $\chi^2$ random variable,
a Gibbs sampler can be easily implemented in this setting
by taking advantage of the corresponding latent variables: $x_i\sim t_{\nu}(\mu,\sigma^2)$
is the marginal of
$$
x_i|V_i,\sigma^2 \sim \mathcal{N}(\mu,V_i\sigma^2) \,,\qquad V_i^{-1}\sim \chi^2_\nu\,.
$$
Once these latent variables are included in the simulation, the conditional posterior distributions of all
parameters are available when using conjugate priors like
$$
\mathbf{p}\sim \mathcal{D}(1,\dots,1)\,,
\quad\mu_j\sim \mathcal{N}(\mu_0,2 \sigma_0^2)\,,%\bar{x},2\sigma^2(x))\,,
\quad\sigma^2_j\sim \mathcal{IG}(\alpha_{\sigma},\beta_{\sigma})\,.
$$
The full conditionals for the Gibbs sampler are a Dirichlet $\mathcal{D}(1+n_1,\dots,1+n_J)$
distribution on the weight vector, inverse Gamma 
$$
\mathcal{IG} \left\{\alpha_{\sigma}+\dfrac{n_j}{2}~,\beta_{\sigma}+\sum_{z_i=j}
\dfrac{(x_i-\mu_j)^2}{2V_i}\right\} 
$$
distributions on the variances $\sigma^2_j$, normal
$$
\mathcal{N} \left(\dfrac{\mu_0\sigma_j^2+2\sigma_0^2\sum_{z_i=j}x_i
V^{-1}_i}{\sigma_j^2+2\sigma_0^2\sum_{z_i=j}V^{-1}_i}
~, \dfrac{2\sigma_0^2\sigma_j^2}{\sigma_j^2+2\sigma_0^2\sum_{z_i=j}V^{-1}_i}
\right) 
$$
distributions on the means $\mu_j$, and inverse Gamma
$$
\mathcal{IG} \left(\dfrac{1}{2}+\dfrac{\nu_j}{2}~,\dfrac{(x_i-\mu_j)^2}{2\sigma^2_j}+\dfrac{\nu_j}{2}\right)
$$
distributions on the $V_i$.

In order to illustrate the performance of the algorithm, we simulated $2,000$ observations 
from the two-component $t$ mixture with $\mu_1=0$, $\mu_2=5$, $\sigma^2_1=\sigma^2_2=1$, 
$\nu_1=5$, $\nu_2=11$ and $p_1=0.3$. The output of the Gibbs sampler is summarized in Figure \ref{Gibbs_SimData}. 
The mixing behaviour of the Gibbs chains seems to be excellent, as they explore neighbourhoods of the true values. \hfill$\blacktriangleleft$
\end{em}
\end{example}

The example below shows that, for specific models and a small number of components, the Gibbs sampler may recover the
symmetry of the target distribution.

\begin{figure}
\begin{center}\includegraphics[width=180pt]{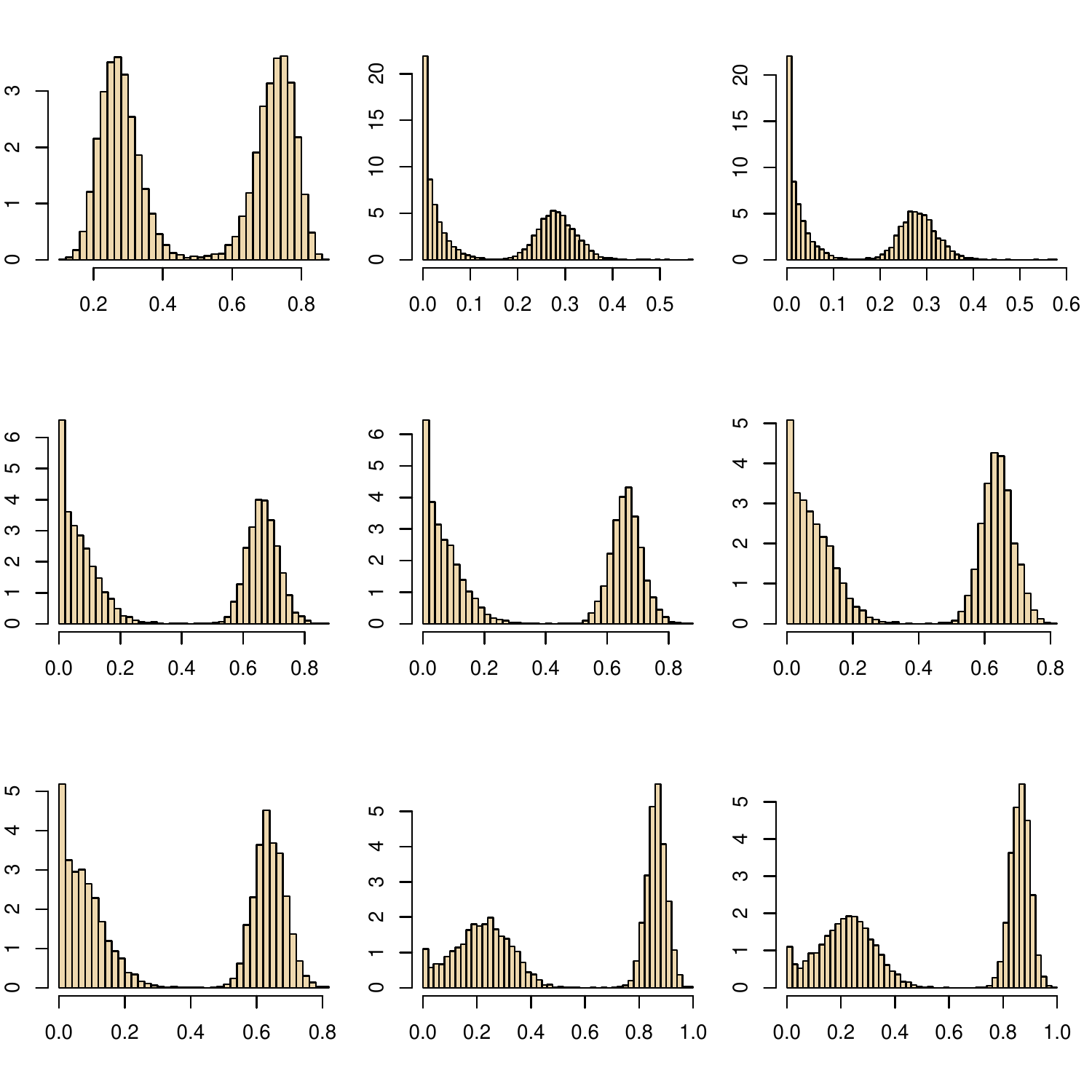}\end{center}
\caption{\label{fig:late}
Latent class model: histograms of $p$ and of the $q_{tj}$'s for $10^4$ iterations of the Gibbs sampler and
the four variables of \cite{stouffer:toby:1951}. The first histogram corresponds to $p$, the
next on the right to $q_{11}$, followed by $q_{21}$ (identical), then $q_{21}$, $q_{22}$, and
so on.
}
\end{figure}

\begin{example} 
\begin{em}
{\bf (Example \ref{exa:stoufftob} continued)} For the latent class model, 
if we use all four variables with two modalities each in \cite{stouffer:toby:1951}, the Gibbs sampler
involves two steps: the completion of the data with the component labels, and the simulation of the probabilities
$p$ and $q_{tj}$ from Beta $\mathcal(B)(s_{tj}+.5,n_j-s_{tj}+.5)$ conditional distributions. For the $216$
observations, the Gibbs sampler seems to converge satisfactorily since the output in Figure \ref{fig:late}
exhibits the perfect symmetry predicted by the theory. We can note that, in this special case, the modes are
well separated, and hence values can be crudely estimated for $q_{1j}$ by a simple graphical identification of
the modes. \hfill$\blacktriangleleft$
\end{em}
\end{example}

\subsection{Metropolis--Hastings approximations}

The Gibbs sampler may fail to escape the attraction of a
local mode, even in a well-behaved case as in Example 1 where the likelihood and the posterior
distributions are bounded and where the parameters are identifiable. Part of the difficulty is
due to the completion scheme that increases the dimension of the simulation space and that reduces
considerably the mobility of the parameter chain.  A standard alternative that does not require
completion and an increase in the dimension is the Metropolis--Hastings algorithm. In fact,
the likelihood of mixture models is available in closed form, being computable in O$(Jn)$ time,
and the posterior distribution is thus available up to a multiplicative constant.

\medskip
{\flushright{\bf {General Metropolis--Hastings algorithm for mixture models}}}
{\sf
\begin{enumerate}
\item[{\sf 0.}]  {\sffamily\bfseries Initialization.}
	Choose $\mathbf{p}^{(0)}$ and $\boldsymbol{\theta}^{(0)}$
\item[{\sf 1.}]  {\sffamily\bfseries Step t.} For $t=1,\ldots$
\begin{enumerate}
\item[{\sf 1.1}] Generate $(\widetilde{\boldsymbol{\theta}},\widetilde{\mathbf{p}})$ from $q\left(\boldsymbol{\theta},\mathbf{p}|
                 \boldsymbol{\theta}^{(t-1)},\mathbf{p}^{(t-1)}\right)$,
\item[{\sf 1.2}] Compute $$r=\frac{f(\mathbf{x}|\widetilde{\boldsymbol{\theta}},\widetilde{\mathbf{p}})\pi(\widetilde{\boldsymbol{\theta}},
                           \widetilde{\mathbf{p}})q(\boldsymbol{\theta}^{(t-1)},\mathbf{p}^{(t-1)}|\widetilde{\boldsymbol{\theta}},
                           \widetilde{\mathbf{p}})}
                           {f(\mathbf{x}|\boldsymbol{\theta}^{(t-1)},\mathbf{p}^{(t-1)})\pi(\boldsymbol{\theta}^{(t-1)},\mathbf{p}^{(t-1)})
                           q(\widetilde{\boldsymbol{\theta}},\widetilde{\mathbf{p}}|\boldsymbol{\theta}^{(t-1)},\mathbf{p}^{(t-1)})},$$ 
\item[{\sf 1.3}] Generate $u\sim\mathscr{U}_{[0,1]}$\\
	If $r>u$ then $(\boldsymbol{\theta}^{(t)},\mathbf{p}^{(t)})=
                 (\widetilde{\boldsymbol{\theta}},\widetilde{\mathbf{p}})$ \\ else $(\boldsymbol{\theta}^{(t)},\mathbf{p}^{(t)})=
                 (\boldsymbol{\theta}^{(t-1)},\mathbf{p}^{(t-1)})$.
\end{enumerate}
\end{enumerate}
}

\medskip
The major difference with the Gibbs sampler is that we need
to choose the proposal distribution $q$, which can be {\em a priori} anything,
and this is a mixed blessing! The most generic
proposal is the random walk Metropolis--Hastings algorithm where each unconstrained 
parameter is the mean of the proposal distribution for the new value, that is,
$$\widetilde{\theta_j}=\theta_j^{(t-1)}+u_j$$ where $u_j\sim\mathcal{N}(0,\zeta^2)$.
However, for constrained parameters like the weights and the variances in a normal mixture
model, this proposal is not efficient.

This is indeed the case for the parameter $\mathbf{p}$, due to the constraint that 
\smash{$\sum_{j=1}^Jp_j\allowbreak=1$.} 
To solve this difficulty, \cite{cappe:robert:ryden:2003} propose to overparameterise the model 
(\ref{eq:mix}) as $$p_j={w_j}\bigg/{\displaystyle \sum_{l=1}^Jw_l}\,,\quad w_j>0\,,$$
thus removing the simulation constraint on the $p_j$'s.
Obviously, the $w_j$'s are not identifiable, but this is not a difficulty from a simulation
point of view and the $p_j$'s remain identifiable (up to a permutation of indices).
Perhaps paradoxically, using overparameterised
representations often helps with the mixing of the corresponding MCMC algorithms 
since they are less constrained by the dataset or the likelihood. The proposed move on the 
$w_j$'s is $\log(\widetilde{w_j})=\log(w_j^{(t-1)})+u_j$ where $u_j\sim\mathcal{N}(0,\zeta^2)$.

\begin{example}
\begin{em}
{\bf (Example \ref{ex:aero} continued)}
%Those degrees of freedom can nonetheless be simulated
%via a Metropolis-Hastings step described in the next section, as illustrated in Figure \ref{Aerosol_t_mixture} which
%provides the estimated densities for four different sets of hyperpriors.
%\begin{figure}[ht!] \begin{center}
%\setlength{\unitlength}{1cm}
%\begin{picture}(10,5.5)
%\put(0,0){\includegraphics[width=10cm, height=5.5cm]{mix_t_aerosol}}
%\end{picture} \end{center} 
%\caption{Histogram of the aerosol particle dataset and $t$ mixtures estimated from four different hyperparameter sets.
%\label{Aerosol_t_mixture}}
%\end{figure}
We now consider the more realistic case when the degrees of freedom of the $t$ distributions are unknown. The Gibbs sampler 
cannot be implemented as such given that the distribution of the $\nu_j$'s is far from standard. A common
alternative \citep{robert:casella:2004} is to introduce a Metropolis step within the Gibbs sampler
to overcome this difficulty. If we use the same Gamma prior distribution with hyperparameters
$(\alpha_{\nu},\beta_{\nu})$ for all the $\nu_j$s, the full conditional density of $\nu_j$ is
$$
\pi(\nu_j|\mathbf{V},\mathbf{z})\propto \left(\dfrac{(\nu_j/2)^{\nu_j/2}}{\Gamma(\nu_j/2)}\right)^{n_j}
\prod_{z_i=j}\dfrac{V_i^{-(\nu_j/2+1)}}{e^{\nu_j/2V_i}} \mathcal{G} (\alpha_{\nu},\beta_{\nu})\,. 
$$
Therefore, we resort to a random walk proposal on the $\log(\nu_j)$'s with scale
$\varsigma=5$. (The hyperparameters are $\alpha_{\nu}=5$ and $\beta_{\nu}=2$.)

In order to illustrate the performances of the algorithm, two cases are considered: (i) all parameters except 
variances ($\sigma^2_1=\sigma^2_2=1$) are unknown and (ii) all parameters are unknown. For a simulated dataset,
the results are given on Figure \ref{MH_fix_sigma_SimData} and Figure \ref{MH_SimData}, respectively. In both cases, the
posterior distributions of the $\nu_j$'s exhibit very large variances, which indicates that the data is very weakly
informative about the degrees of freedom. The Gibbs sampler does not mix well-enough to recover the symmetry in the marginal approximations.
The comparison between the estimated densities for both cases with the setting is given in Figure \ref{SimData}.
The estimated mixture densities are indistinguishable and the fit to the simulated dataset is quite adequate.
Clearly, the corresponding Gibbs samplers have recovered
correctly one and only one of the $2$ symetric modes. 

\begin{figure}[ht!] \begin{center}
\setlength{\unitlength}{1cm}
\begin{picture}(12,4.5)
\put(0,0){\includegraphics[width=12cm, height=4.5cm]{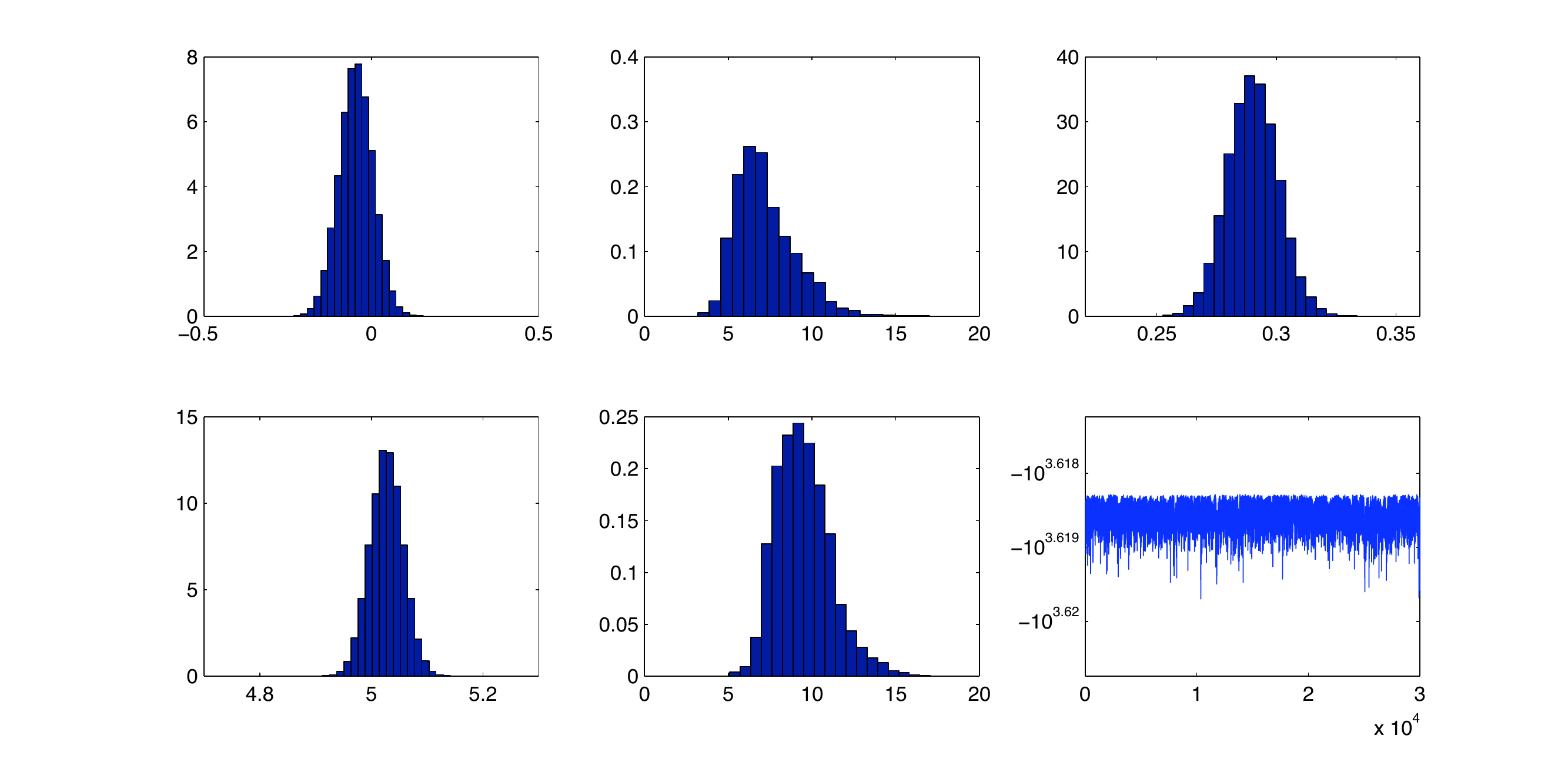}}
\end{picture} \end{center} 
\caption{Histograms of the parameters $\mu_1,\nu_1, p_1,\mu_2,\nu_2$ when the variance parameters are
known, and evolution of the log-likelihood for a simulated $t$ mixture with $2,000$ points, based on $3\times 10^4$ MCMC iterations.
\label{MH_fix_sigma_SimData}}
\end{figure}

\begin{figure}[ht!] \begin{center}
\setlength{\unitlength}{1cm}
\begin{picture}(12,5.5)
\put(-1,0){\includegraphics[width=14cm, height=5.5cm]{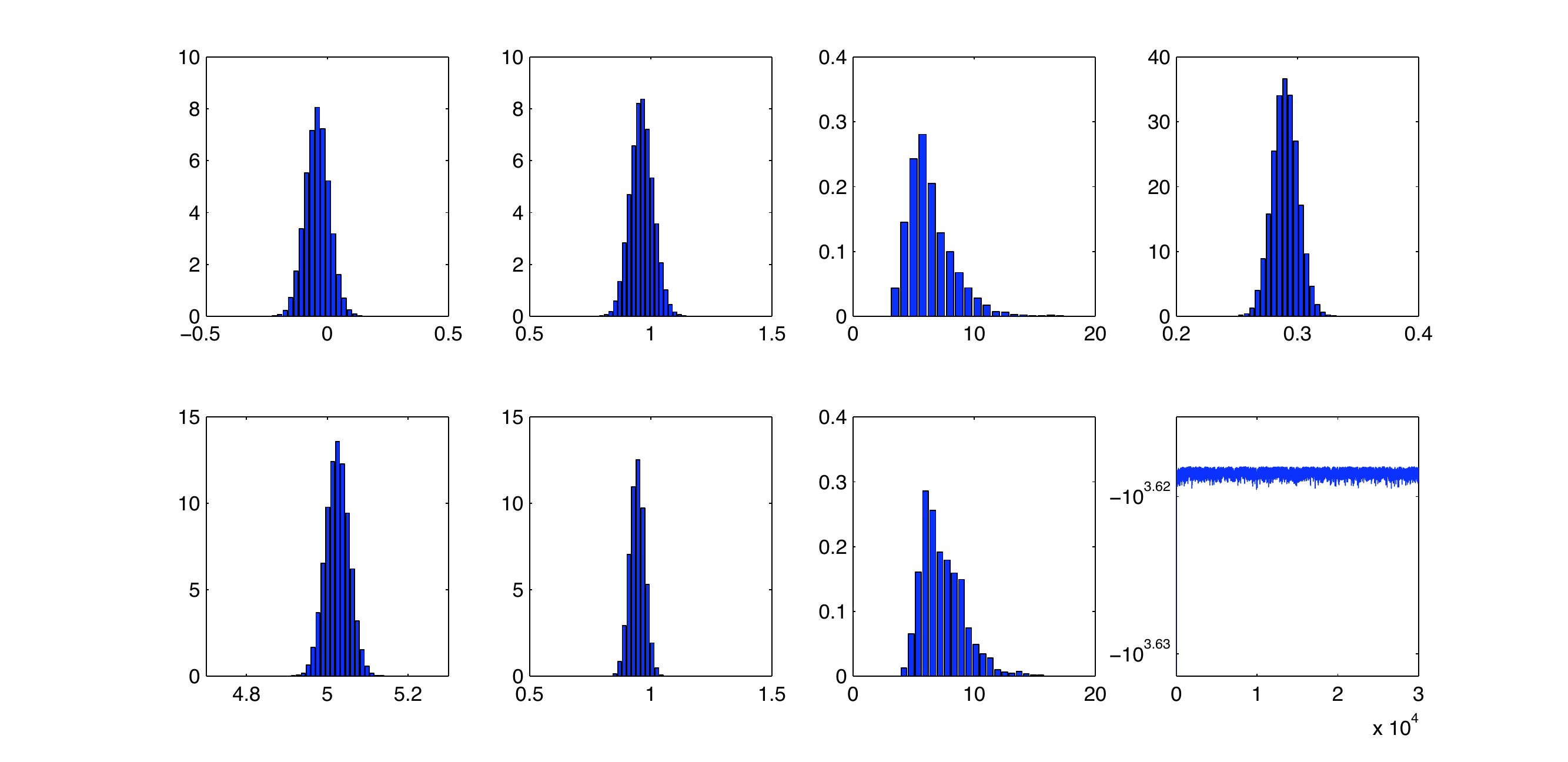}}
\end{picture} \end{center} 
\caption{Histograms of the parameters $\mu_1,\sigma_1,\nu_1,p_1,\mu_2,\sigma_2,\nu_2$,
and evolution of the log-likelihood for a simulated $t$ mixture with $2,000$ points, based on $3\times 10^4$ MCMC
iterations.
\label{MH_SimData}}
\end{figure}

\begin{figure}[ht!] \begin{center}
\setlength{\unitlength}{1cm}
\begin{picture}(8,4.5)
\put(0,0){\includegraphics[width=8cm, height=4.5cm]{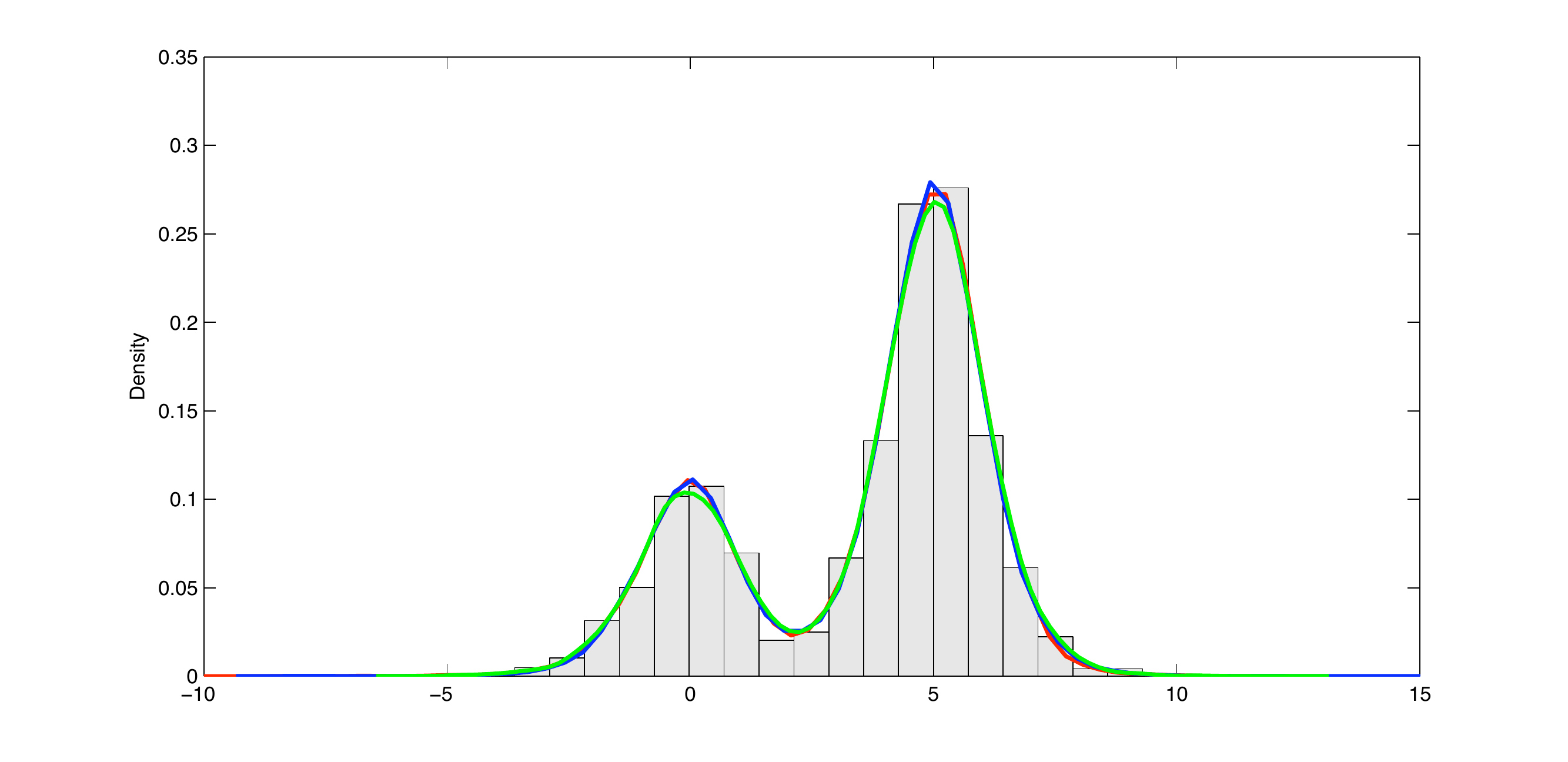}}
\end{picture} \end{center} 
\caption{Histogram of the simulated dataset, compared 
with estimated $t$ mixtures with known $\sigma^2$ {\em (red)}, 
known $\nu$ {\em (green)}, and when all parameters are unknown {\em (blue)}.
\label{SimData}}
\end{figure}

We now consider the aerosol particle dataset described in Example \ref{ex:aero}. We use the same prior distributions on the $\nu_j$'s as before, that is $\mathcal{G}(5,2)$.
Figure \ref{MH_AerosolData} summarises the output of the MCMC algorithm.
Since there is no label switching and only two components, we choose to estimate the parameters by the empirical
averages, as illustrated in Table \ref{table 1}. As shown by Figure \ref{fig:aero1},
both $t$ mixtures and normal mixtures fit the aerosol data reasonably well. \hfill$\blacktriangleleft$
\end{em}
\end{example}

\begin{figure}[ht!] \begin{center}
\setlength{\unitlength}{1cm}
\begin{picture}(12,5.5)
\put(-2.5,0){\includegraphics[width=16cm, height=5.5cm]{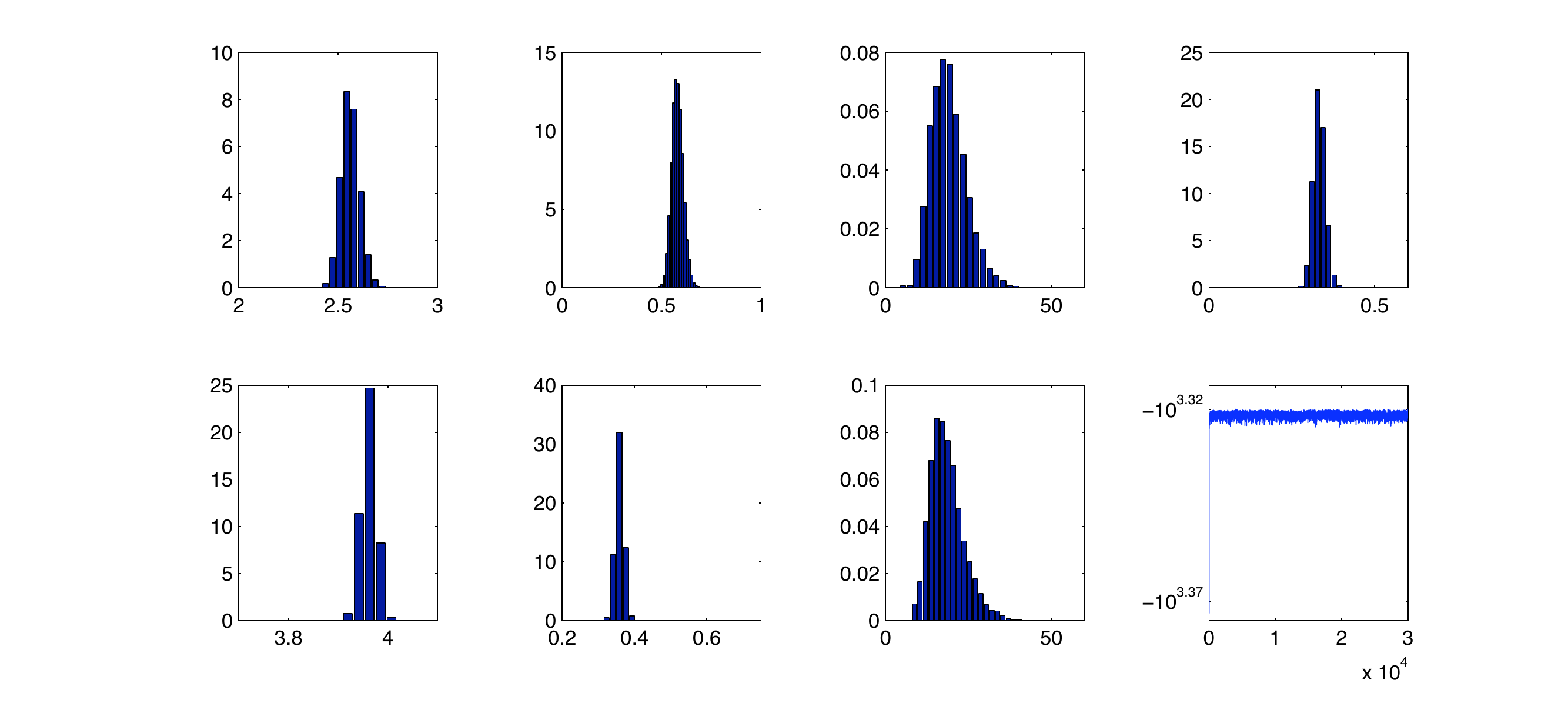}}
\end{picture} \end{center} 
\caption{Histograms of parameters ($\mu_1,\sigma_1, \nu_1, p_1,\mu_1,\sigma_2,\nu_2$) and log-likelihood of a mixture of $t$ distributions based on $30,000$ iterations
and the aerosol data.
\label{MH_AerosolData}}
\end{figure}

\begin{table}\begin{center}\begin{tabular}{c c c c c c c c c}
\hline  &$\mu_1$&$\mu_2$&$\sigma_1$&$\sigma_2$&$\nu_1$&$\nu_2$&$p_1$ \\ \hline
{\rm Student}&2.5624&3.9918&0.5795&0.3595&18.5736&19.3001&0.3336 \\
\hline 
{\rm Normal}&2.5729&3.9680&0.6004&0.3704&-&-&0.3391\\
\hline 
\end{tabular} 
\caption{Estimates of the parameters for the aerosol dataset compared for $t$ and normal mixtures.
\label{table 1}}\end{center}\end{table}

%%%%%%%%%%%%%%%%%%%%%%%%%%%%%%%%%%%%%%%
\section{Inference for mixture models with an unknown number of components}
\label{sec:mocho}
%%%%%%%%%%%%%%%%%%%%%%%%%%%%%%%%%%%%%%%

Estimation of $J$, the number of components in (\ref{eq:mix}), is a special type of model choice problem, for which there is
a number of possible solutions:
\begin{enumerate}\renewcommand{\theenumi}{(\roman{enumi})}
\item[(i)] direct computation of the Bayes factors \citep{Kass:Raftery:1995,chib:1995};
\item[(ii)] evaluation of an entropy distance \citep{Mengersen:Robert:1996,Sahu:Cheng:2003};
\item[(iii)] generation from a joint distribution across models via
reversible jump MCMC \citep{Richardson:Green:1997} or via birth-and-death processes \citep{Stephens:2000} ;
\end{enumerate}
depending on whether the perspective is on testing or estimation. 
We refer to \cite{marin:mengersen:robert:2005} for a short description of
the reversible jump MCMC solution, a longer survey being available in \cite{robert:casella:2004} and a specific
description for mixtures--including an {\sf R} package---being provided in 
\cite{marin:robert:2007}. The alternative birth-and-death processes proposed in \cite{Stephens:2000} has not generated as much follow-up,
except for \cite{cappe:robert:ryden:2003} who showed that the essential mechanism in this approach was the same as with
reversible jump MCMC algorithms.

We focus here on the first two approaches, because,
first, the description of reversible jump MCMC algorithms require much care and therefore more space than
we can allow to this paper and, second, this description exemplifies recent advances in the derivation of Bayes factors.
These solutions pertain more strongly to the testing perspective, the
entropy distance approach being based on the Kullback--Leibler divergence between a
$J$ component mixture and its projection on the set of $J-1$ mixtures, in the same
spirit as in \cite{Dupuis:Robert:2003}. Given that the calibration of the Kullback divergence
is open to various interpretations \citep{Mengersen:Robert:1996,goutis:robert:1998,Dupuis:Robert:2003},
we will only cover here some proposals regarding approximations of the Bayes factor oriented towards the 
direct exploitation of outputs from single model MCMC runs. 

In fact, the major difference between approximations of Bayes factors based on those outputs and approximations
based on the output from the reversible jump chains is that the latter requires a sufficiently efficient choice of proposals
to move around models, which can be difficult despite significant recent advances
\citep{brooks:giudici:roberts:2003}. If we can instead concentrate the simulation effort on single models,
the complexity of the algorithm decreases (a lot) and there exist ways to evaluate the performance of the
corresponding MCMC samples. In addition, it is often the case that few models are in competition when
estimating $J$ and it is therefore possible to visit the whole range of potentials models in an exhaustive manner.

We have
$$
f_J(\bx|\boldsymbol{\lambda}_J)=\prod_{i=1}^n\sum_{j=1}^Jp_jf(x_i|\theta_j)
$$
where $\boldsymbol{\lambda}_J=\left(\boldsymbol{\theta},\boldsymbol{p}\right)=\left(\theta_1,\ldots,\theta_J,p_1,\ldots,p_J\right)$.
Most solutions \citep[see, e.g.][Section 5.4]{fruhwirth:2006}
revolve around an importance sampling approximation to the marginal likelihood integral
$$
m_J(x) = \int f_J(\bx|\boldsymbol{\lambda}_J)\,\pi_J(\boldsymbol{\lambda}_J)\,\dd \boldsymbol{\lambda}_J
$$
where $J$ denotes the model index (that is the number of components in the present case). For instance, \cite{liang:wong:2001} use bridge sampling
with simulated annealing scenarios to overcome the label switching problem. \cite{steele:raftery;emond:2006}
rely on defensive sampling and the use of conjugate priors to reduce the integration to the space of latent
variables (as in \citealp{Casella:Robert:Wells:1999}) with an iterative construction of the importance function.
\cite{fruhwirth:2004} also centers her approximation of the marginal likelihood on a bridge sampling strategy,
with particular attention paid to identifiability constraints.
A different possibility is to use \cite{gelfand:dey:1994} representation:
starting from an arbitrary density $g_J$, the equality
\begin{align*}
1 &= \int g_J(\boldsymbol{\lambda}_J) \,\dd \boldsymbol{\lambda}_J = \int \frac{g_J(\boldsymbol{\lambda}_J)}{f_J(\bx|\boldsymbol{\lambda}_J)\,\pi_J(\boldsymbol{\lambda}_J)} 
f_J(\bx|\boldsymbol{\lambda}_J)\,\pi_J(\boldsymbol{\lambda}_J) \,\dd \boldsymbol{\lambda}_J \\ &= m_J(\bx)
\int \frac{g_J(\boldsymbol{\lambda}_J)}{f_J(\bx|\boldsymbol{\lambda}_J)\,\pi_J(\boldsymbol{\lambda}_J)} \pi_J(\boldsymbol{\lambda}_J|\bx) \,\dd \boldsymbol{\lambda}_J
\end{align*}
implies that a potential estimate of $m_J(\bx)$ is
$$
\hat m_J(\bx) = 1 \bigg/ \frac{1}{T} \sum_{t=1}^T 
\frac{g_J(\boldsymbol{\lambda}_J^{(t)})}{f_J(\bx|\boldsymbol{\lambda}_J^{(t)})\,\pi_J(\boldsymbol{\lambda}_J^{(t)})}
$$
when the $\boldsymbol{\lambda}_J^{(t)}$'s are produced by a Monte Carlo or an MCMC sampler targeted at $\pi_J(\boldsymbol{\lambda}_J|\bx)$.
While this solution can be easily implemented in low dimensional settings
\citep{chopin:robert:2007}, calibrating the auxiliary density
$g_k$ is always an issue. The auxiliary density could be selected as a non-parametric estimate of $\pi_k(\boldsymbol{\lambda}_J|x)$
based on the sample itself but this is very costly. Another difficulty is that the estimate may have an infinite
variance and thus be too variable to be trustworthy, as experimented by \cite{fruhwirth:2004}. 

Yet another approximation to the integral $m_J(\bx)$ is to consider
it as the expectation of $f_J(\bx|\boldsymbol{\lambda}_J)$, when $\boldsymbol{\lambda}_J$ is distributed from the prior. While
a brute force approach simulating $\boldsymbol{\lambda}_J$ from the prior distribution is requiring a huge number
of simulations \citep{neal:1999}, a Riemann based alternative is proposed by \cite{skilling:2007a} under
the denomination of {\em nested sampling}; however, \cite{chopin:robert:2007} have shown in the case of mixtures
that this technique could lead to uncertainties about the quality of the approximation.

We consider here a further solution, first proposed by \cite{chib:1995}, that is straightforward
to implement in the setting of mixtures (see \citealp{chib:jeliazkov:2001} for extensions).
Although it came under criticism by \cite{neal:1999} (see also \citealp{fruhwirth:2004}), 
we show below how the drawback pointed by the latter can easily be removed. Chib's
(\citeyear{chib:1995})  method is directly based on the expression of the marginal
distribution (loosely called {\em marginal likelihood} in this section) in Bayes' theorem:
$$
m_J(\bx) = \frac{f_J(\bx|\boldsymbol{\lambda}_J)\,\pi_J(\boldsymbol{\lambda}_J)}{\pi_J(\boldsymbol{\lambda}_J|\bx)}
$$
and on the property that the rhs of this equation is constant in $\boldsymbol{\lambda}_J$. Therefore, if an
arbitrary value of $\boldsymbol{\lambda}_J$, $\boldsymbol{\lambda}_J^*$ say, is selected and if a good approximation to $\pi_J(\boldsymbol{\lambda}_J|\bx)$
can be constructed, $\hat{\pi}_J(\boldsymbol{\lambda}_J|\bx)$, Chib's (\citeyear{chib:1995})
approximation to the marginal likelihood is
\begin{equation}\label{eq:chib}
\hat{m}_J(\bx) = \frac{f_J(\bx|\boldsymbol{\lambda}_J^*)\,\pi_J(\boldsymbol{\lambda}_J^*)}{\hat{\pi}_J(\boldsymbol{\lambda}_J^*|\bx)}\,.
\end{equation}
In the case of mixtures, a natural approximation to $\pi_J(\boldsymbol{\lambda}_J|\bx)$ is the Rao-Blackwell estimate
$$
\hat{\pi}_J(\boldsymbol{\lambda}_J^*|\bx) = \frac{1}{T}\,\sum_{t=1}^T \pi_J(\boldsymbol{\lambda}_J^*|\bx,\bz^{(t)})\,,
$$
where the $\bz^{(t)}$'s are the latent variables simulated by the MCMC sampler.
To be efficient, this method requires 
\begin{itemize}
\item[(a)] a good choice of $\boldsymbol{\lambda}_J^*$ but, since in the case of
mixtures, the likelihood is computable, $\boldsymbol{\lambda}_J^*$ can be chosen as the MCMC approximation to
the MAP estimator and, 
\item[(b)] a good approximation to $\pi_J(\boldsymbol{\lambda}_J|\bx)$. 
\end{itemize}
This later requirement is the core of Neal's (\citeyear{neal:1999}) criticism: while, at a formal level,
$\hat{\pi}_J(\boldsymbol{\lambda}_J^*|\bx)$ is a converging (parametric) approximation to $\pi_J(\boldsymbol{\lambda}_J|\bx)$
by virtue of the ergodic theorem, this obviously requires the chain $(\bz^{(t)})$ to converge to its
stationarity distribution. Unfortunately, as discussed previously, in the case of mixtures, the Gibbs sampler rarely converges
because of the label switching phenomenon described in Section \ref{sub:ifaible}, so the
approximation $\hat{\pi}_J(\boldsymbol{\lambda}_J^*|\bx)$ is untrustworthy. \cite{neal:1999} demonstrated via
a numerical experiment that \eqref{eq:chib} is significantly different from the true value $m_J(\bx)$ when
label switching does not occur. There is, however, a fix to this problem, also explored by
\cite{berkhof:vanmechelen:gelman:2003}, which is to recover the
label switching symmetry a posteriori, replacing $\hat{\pi}_J(\boldsymbol{\lambda}_J^*|\bx)$ in \eqref{eq:chib} above with
$$
\hat{\pi}_J(\boldsymbol{\lambda}_J^*|\bx) = \frac{1}{T\,J!}\,
\sum_{\sigma\in\mathfrak{S}_J}\sum_{t=1}^T \pi_J(\sigma(\boldsymbol{\lambda}_J^*)|\bx,\bz^{(t)})\,,
$$
where $\mathfrak{S}_J$ denotes the set of all permutations of $\{1,\ldots,J\}$ and
$\sigma(\boldsymbol{\lambda}_J^*)$ denotes the transform of $\boldsymbol{\lambda}_J^*$ where components are
switched according to the permutation $\sigma$. Note that the permutation can
equally be applied to $\boldsymbol{\lambda}_J^*$ or to the $\bz^{(t)}$'s but that the former
is usually more efficient from a computational point of view given that the sufficient statistics only
have to be computed once.
The justification for this modification either stems from a Rao-Blackwellisation argument,
namely that the permutations are ancillary for the problem and should be integrated out, or
follows from the general framework of \cite{kong:mccullagh:nicolae:tan:meng:2003} where
symmetries in the dominating measure should be exploited towards the improvement of the variance of Monte
Carlo estimators.

\begin{example} 
\begin{em}
{\bf (Example \ref{ex:galax} continued)} In the case of the
normal mixture case and the galaxy dataset, using Gibbs sampling,
label switching does not occur. If we compute $\log \hat{m}_J(\bx)$ using only the original
estimate of \cite{chib:1995} (\ref{eq:chib}), the [logarithm of the] estimated marginal likelihood is $\hat\rho_J(\bx) = -105.1396$
for $J=3$ (based on $10^3$ simulations), while introducing the permutations leads to
$\hat\rho_J(\bx) = -103.3479$. As already noted by \cite{neal:1999}, the difference between
the original Chib's (\citeyear{chib:1995}) approximation and the true marginal likelihood is
close to $\log(J!)$ (only) when the Gibbs sampler remains concentrated around a single mode of the
posterior distribution. In the current case, we have that $-116.3747+\log(2!)=-115.6816$ exactly!
(We also checked this numerical value against a brute-force estimate obtained by simulating from
the prior and averaging the likelihood, up to fourth digit agreement.)
A similar result holds for $J=3$, with $-105.1396+\log(3!)=-103.3479$. Both \cite{neal:1999} 
and \cite{fruhwirth:2004} also
pointed out that the $\log(J!)$ difference was unlikely to hold for larger values of $J$ as the
modes became less separated on the posterior surface and thus the Gibbs sampler was more
likely to explore incompletely several modes. For $J=4$, we get for instance that the original
Chib's (\citeyear{chib:1995}) approximation is $-104.1936$, while the average over permutations
gives $-102.6642$. Similarly, for $J=5$, the difference between $-103.91$ and $-101.93$ is less
than $\log(5!)$. The $\log(J!)$ difference cannot therefore be used as a direct correction for Chib's 
(\citeyear{chib:1995}) approximation because of this difficulty in controlling the amount of overlap.
However, it is unnecessary since using the permutation average resolves the difficulty. Table
\ref{tab:galax} shows that the prefered value of $J$ for the galaxy dataset and the current choice
of prior distribution is $J=5$. \hfill$\blacktriangleleft$
\end{em}
\end{example}

\begin{table}\begin{center}
\begin{tabular}{l|ccccccc}
J &2 &3 &4 &5 &6 &7 &8\\
\hline
$\hat\rho_J(\bx)$ &-115.68 &-103.35 &-102.66 &-101.93 &-102.88 &-105.48 &-108.44\\
\end{tabular}\end{center}
\caption{\label{tab:galax}
Estimations of the marginal likelihoods by the symmetrised Chib's approximation (based on $10^5$
Gibbs iterations and, for $J>5$, $100$ permutations selected at random in $\mathfrak{S}_J$).
}
\end{table}

When the number of components $J$ grows too large for all permutations in $\mathfrak{S}_J$
to be considered in the average, a (random) subsample of permutations can be simulated to keep 
the computing time to a reasonable level when keeping the identity as one of the permutations, as in
Table \ref{tab:galax} for $J=6,7$.
% (in the case of Example \ref{ex:galaxib}, this occurred for $k=8$). 
(See \citealp{berkhof:vanmechelen:gelman:2003} for another solution.)
Note also that the discrepancy between the original Chib's (\citeyear{chib:1995}) approximation
and the average over permutations is a good indicator of the mixing properties of the Markov 
chain, if a further convergence indicator is requested.

\begin{example}
\begin{em}
{\bf (Example \ref{exa:stoufftob} continued)}
For instance, in the setting of Example \ref{exa:stoufftob} with $a=b=1$, both the approximation of \cite{chib:1995}
and the symmetrized one are identical. When comparing a single class model with a two class model,
the corresponding (log-)marginals are 
$$
\hat\rho_1(\bx) = \prod_{i=1}^4 \frac{\Gamma(1)}{\Gamma(1/2)^2}\,
\frac{\Gamma(n_i+1/2)\Gamma(n-n_i+1/2)}{\Gamma(n+1)}
=-552.0402
$$
and $\hat\rho_2(\bx)\approx -523.2978$, giving a clear preference to the two class model. \hfill$\blacktriangleleft$
\end{em}
\end{example}

\section*{Acknowledgements}
We are grateful to the editors for the invitation as well as to Gilles Celeux for a careful reading of an earlier draft
and for important suggestions related with the latent class model.

\small
% \bibliographystyle{apalike}
% \bibliography{biblio}

\end{document}